\documentclass[cits]{JINST}
\usepackage{epsfig}
\usepackage{subfigure}

\title{A Comparative Numerical Study on GEM, MHSP and MSGC}
\author{
Purba Bhattacharya\thanks{Corresponding author},
Supratik Mukhopadhyay,
Nayana Majumdar,
Sudeb Bhattacharya\\
Applied Nuclear Physics Division, Saha Institute of Nuclear Physics,\\
 1/AF, Bidhannagar, Kolkata 700064, India\\
\\
E-mail: \email{purba.bhattacharya@saha.ac.in}}

\abstract{In this work, we have tried to develop a detailed understanding of the physical processes
occurring in those variants of Micro Pattern Gas Detectors (MPGDs) that share micro hole and micro strip
geometry, like GEM, MHSP and MSGC etc. Some of the important and
fundamental characteristics of these detectors such as gain, transparency, efficiency and their
operational dependence on different device parameters have been estimated following detailed numerical
simulation of the detector dynamics. We have used a relatively new simulation framework developed especially
for the MPGDs that combines packages such as GARFIELD, neBEM, MAGBOLTZ and HEED.
The results compare closely with the available
experimental data. This suggests the efficacy of the framework to model the intricacies
of these micro-structured detectors in addition to providing insight into their inherent complex dynamical
processes.}

\keywords{Gaseous detectors; Detector modelling and simulations II (electric fields, charge transport, 
multiplication and induction, pulse formation, electron emission, etc); Micropattern gaseous 
detectors(MSGC, GEM, THGEM, RETHGEM, MHSP, MICROPIC, MICROMEGAS, InGrid, etc)}

\begin{document}

\section{Introduction}
\label{sec:intro}
Micro Pattern Gas Detectors (MPGDs) \cite{MPGD}, a recent addition to the gas detector
family that utilize semiconductor fabrication techniques, have found wide
applications in different
experiments involving astro-particle physics, high energy physics,
rare event detection, radiation imaging etc.
Despite the widespread acceptance of
MPGDs, a thorough understanding of their working principle is yet to be achieved. 
 
In this paper we have used numerical simulation \cite{Simulation} as a tool of exploration to evaluate
fundamental features of a MHSP detector \cite{VelosoMHSP1}. In the process we have also simulated 
a GEM \cite{SauliGEM} and a MSGC \cite{Oed}, having similar geometrical and material features.
A comprehensive comparison of their characteristics, achieved through the
design variation among these detectors, have been presented.
The study includes extensive computation of electrostatic field configuration within
a given device and its variation for different voltage settings. 
Some of the fundamental properties
like gain, collection efficiency have been estimated too, although of a preliminary nature
at present, and compared to the reported experimental
results.
 
We have used
the recently developed simulation framework Garfield \cite{Garfield1, Garfield2} that combines
packages such as neBEM \cite{neBEM1, neBEM2, neBEM3, neBEM4}, Magboltz \cite{Magboltz1, Magboltz2}
and Heed \cite{HEED1, HEED2}. It may be mentioned here that the simulation framework used in this 
work was augmented in 2009 through the addition of the neBEM toolkit to carry out
3D electrostatic field simulation.
Erlier,
Garfield had to import field-maps from one of the several commercial FEM packages in order to study 3D gas detectors.
Due to the exact foundation expressions based on the Green's functions, the neBEM approach has been found to be exceptionally accurate in the complete physical domain, including the near field.
This fact, in addition to other generic advantages of BEM over FEM, makes neBEM a strong candidate as a field-solver for MPGD related computations.
Some of the major generic advantages of BEM  are its ability to estimate the
field directly and to handle open geometries.
FEM also has several advantages to its credit, such as flexibility, ability to
simulate non-linear problems and huge popularity.
While the Garfield + neBEM framework has been applied for modelling
Micromegas detectors on several occasions \cite{Micromegas, Purba}, very few reports
are available for detectors having a fair amount
of dielectric material, such as MSGC, GEM or MHSP.

\section{Geometry Modelling}
\label{sec:model}

\begin{figure}[hbt]
\centering
\subfigure[]
{\label{schematic}\includegraphics[height=0.3\textheight]{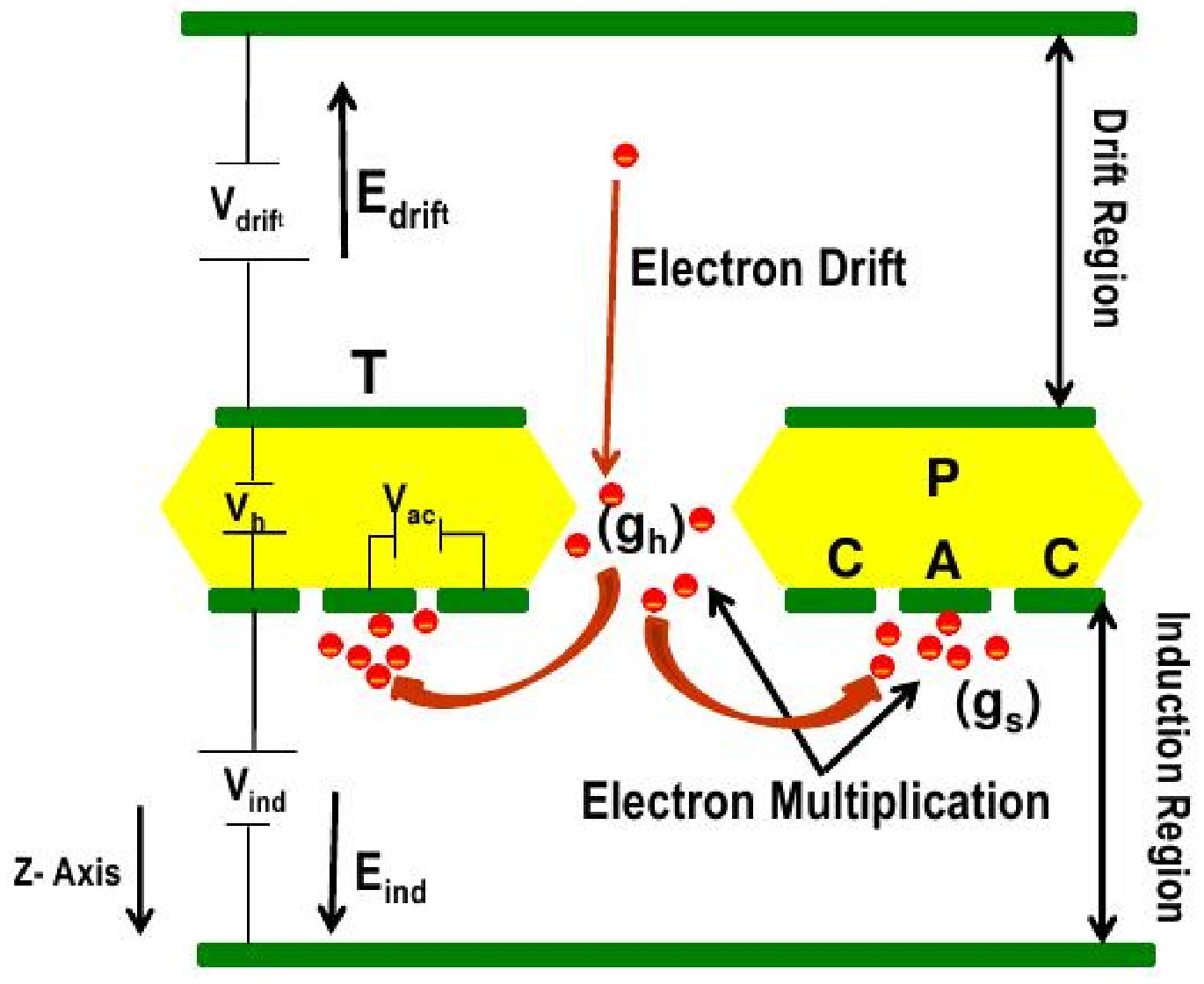}}\\
\subfigure[]
{\label{MHSPArea3D}\includegraphics[height=0.125\textheight]{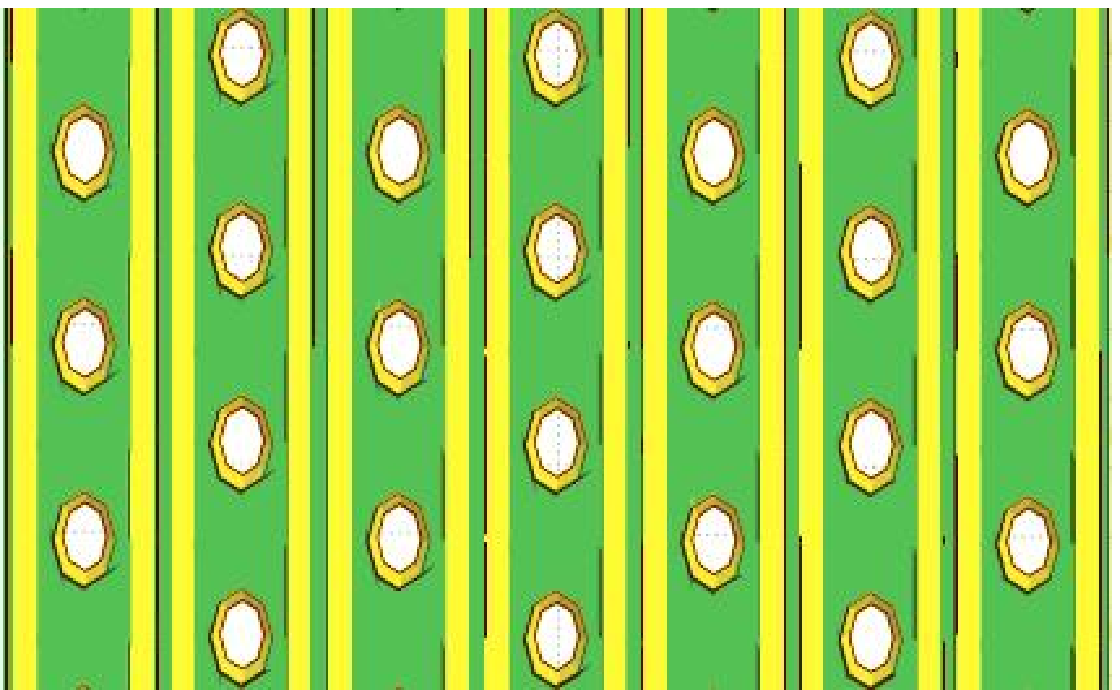}}
\subfigure[]
{\label{GEMArea3D}\includegraphics[height=0.125\textheight]{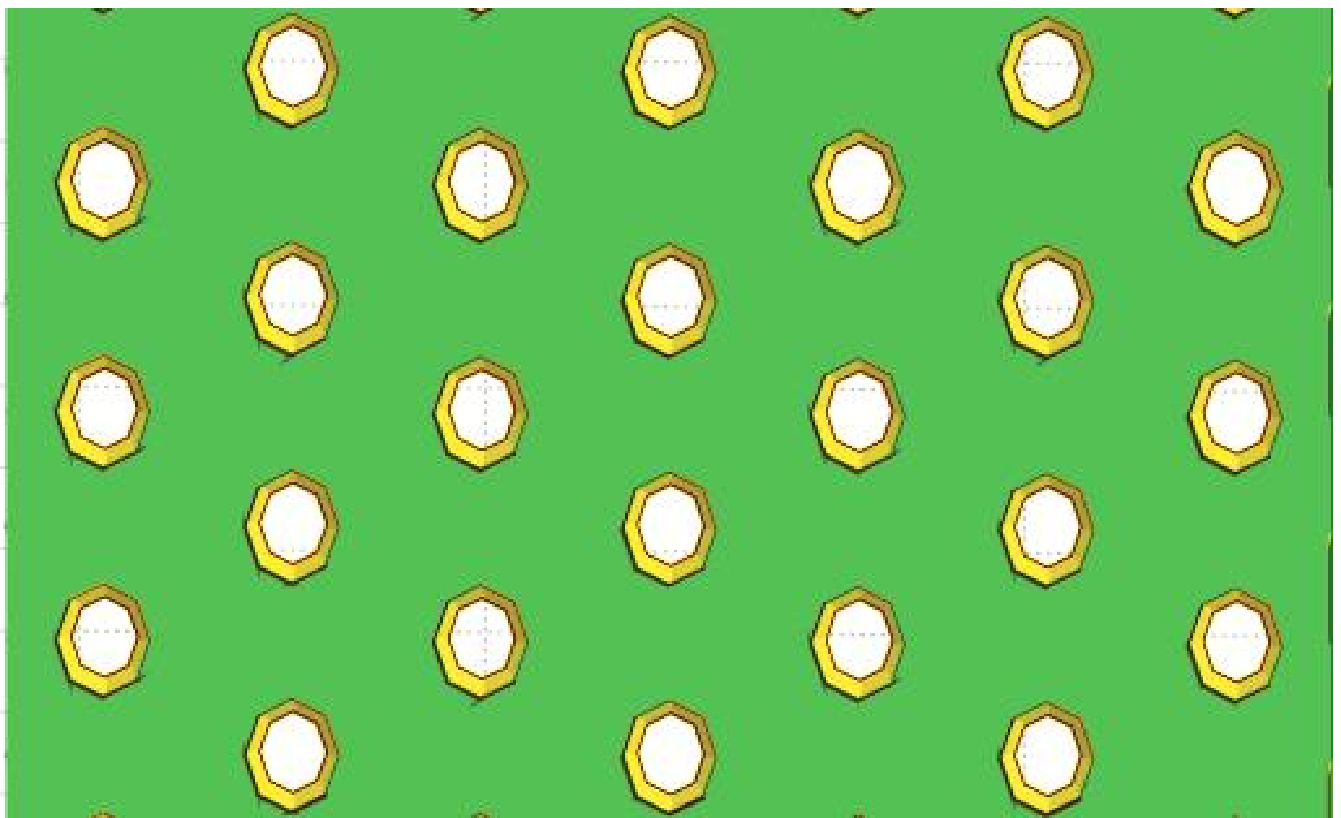}}
\subfigure[]
{\label{MSGCArea3D}\includegraphics[height=0.125\textheight]{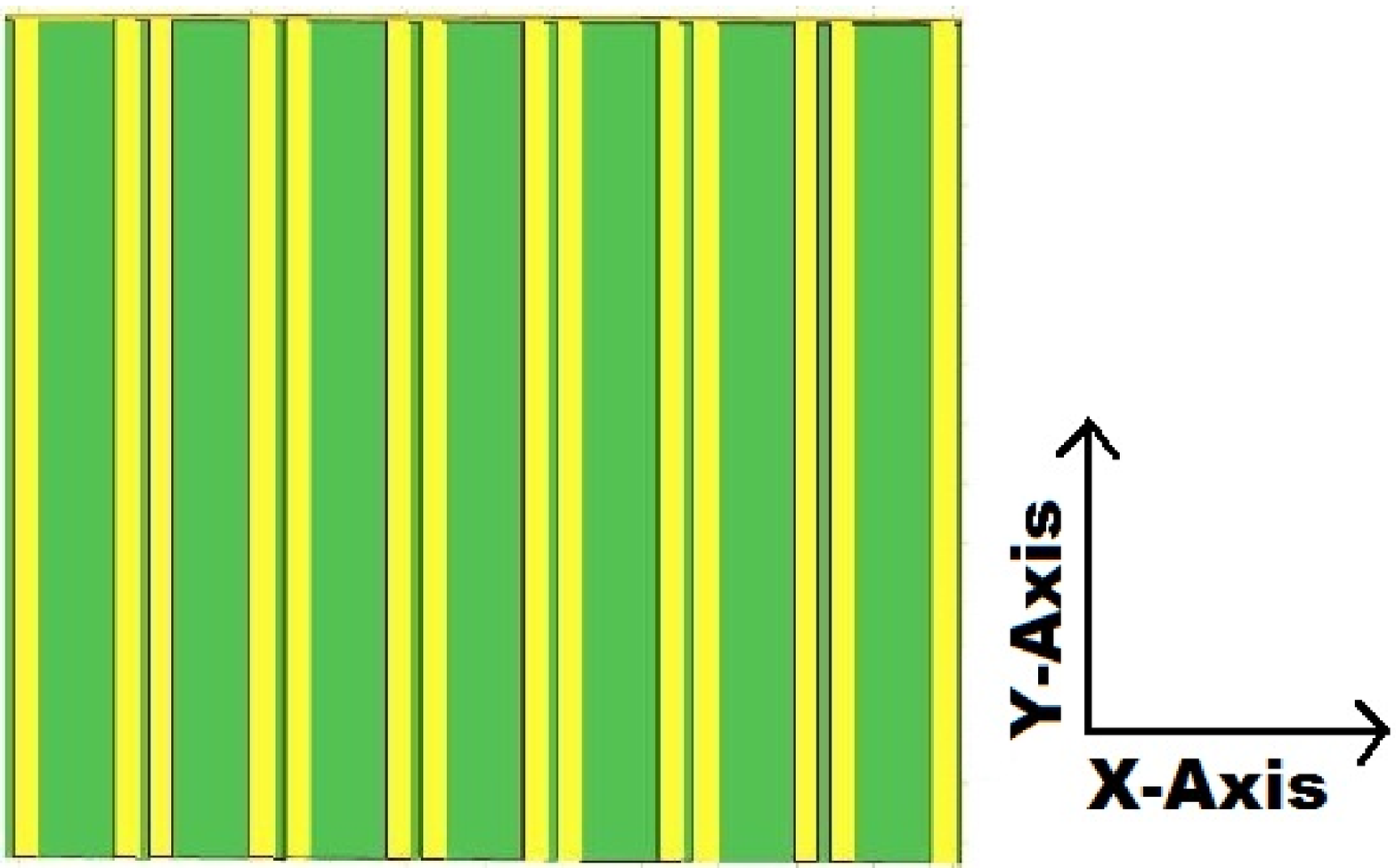}}
\caption{(a) Schematic representation for MHSP, (b) Lower surface of MHSP 
(c) GEM, (d) MSGC}
\label{Area}
\end{figure}

A schematic drawing to represent a MHSP has been depicted in the figure \ref{schematic}.
Basically a MHSP merges the MSGC and the GEM features in a single, double-sided element. 
The top surface looks similar to a GEM (figure \ref{GEMArea3D}).
The bottom plane (figure \ref{MHSPArea3D}) is etched for 
parallel anode (A) and cathode (C) strips (quite similar to a MSGC shown in figure \ref{MSGCArea3D}) with 
holes within the cathode strips (unlike in a MSGC where there are no holes).
The anode and cathode strips
in the MHSP (also for the MSGC) are kept at a potential difference, $\mathrm{V}_\mathrm{ac}$, 
while that across a hole between the top grid (denoted
by T) and the cathode at the bottom is $\mathrm{V}_\mathrm{h}$ for both MHSP and GEM.
Figure \ref{schematic} also depicts the potential difference applied in the drift
region, $\mathrm{V}_\mathrm{drift}$.
Similarly,
a potential difference in the induction region, $\mathrm{V}_\mathrm{ind}$ is also maintained. 
Corresponding electric fields
are denoted by $\mathrm{E}_\mathrm{drift}$ and $\mathrm{E}_\mathrm{ind}$, respectively.
It may be mentioned here that for the MSGC there is no induction region.
Two stages of amplification of electrons in a MHSP
have been denoted in figure \ref{schematic} as $\mathrm{g}_\mathrm{h}$ and 
$\mathrm{g}_\mathrm{s}$ which represent the multiplication inside the hole and that
near the strips, respectively.
The design parameters, considered in the numerical work, are mentioned in table \ref{design}.

\begin{table}[h]
\caption{Design Parameters}\label{design}
\begin{center}
\begin{tabular}{|c|c|c|c|}
\hline
& MHSP & GEM & MSGC \\
& (\cite{VelosoMHSP2}) & (\cite{VelosoMHSP2, BachmannGEM, SharmaGEM}) & (similar to MHSP) \\
\hline
Polymer substrate (P) thickness & $50~\mu\mathrm{m}$ & $50~\mu\mathrm{m}$ & $50~\mu\mathrm{m}$ \\
\hline
copper coating thickness & $5~\mu\mathrm{m}$ & $5~\mu\mathrm{m}$ & $5~\mu\mathrm{m}$ \\
\hline
hole diameter in the copper layer & 70 $\mu\mathrm{m}$ & 70 $\mu\mathrm{m}$ & \\
\hline
hole diameter at the middle of the Polymer substrate & 50 $\mu\mathrm{m}$ & 50 $\mu\mathrm{m}$ & \\
\hline
hole to hole pitch (Y direction) & $140~\mu\mathrm{m}$ & $140~\mu\mathrm{m}$ & \\
\hline
anode width & $15~\mu\mathrm{m}$ & & $15~\mu\mathrm{m}$ \\
\hline 
cathode width & $100~\mu\mathrm{m}$ & & $100~\mu\mathrm{m}$ \\
\hline
anode to cathode gap (edge to edge) & $30~\mu\mathrm{m}$ & & $30~\mu\mathrm{m}$ \\
\hline
anode to anode pitch & $175~\mu\mathrm{m}$ & & $175~\mu\mathrm{m}$ \\
\hline
\end{tabular}
\end{center}
\end{table}

\section{Results and Discussions}
\label{results}
\subsection{Electrostatic Configuration}
\label{sec:field}

\begin{figure}[hbt]
\centering
\subfigure[]
{\label{FieldMHSPGEMCenter}\includegraphics[height=0.2\textheight]{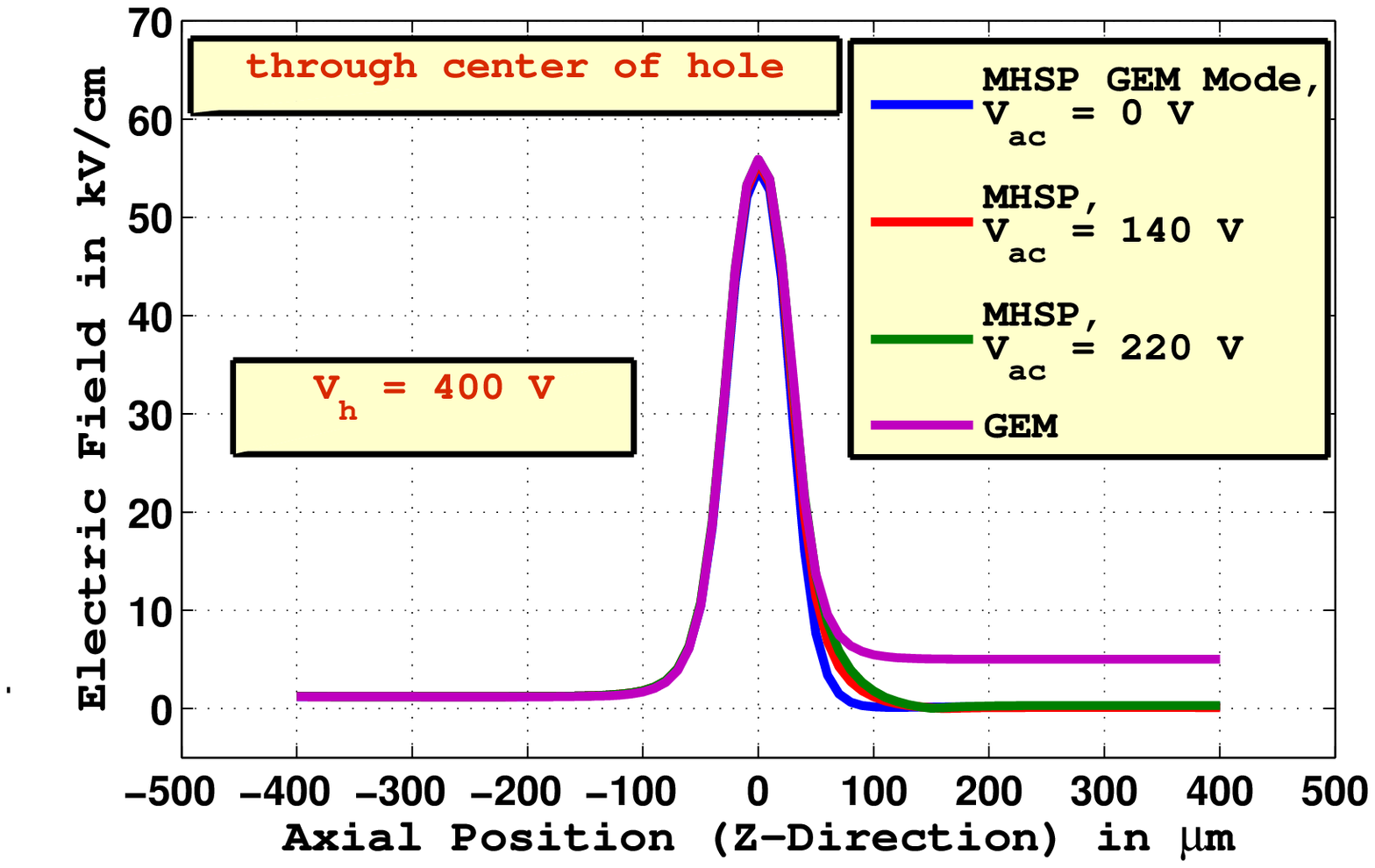}}
\subfigure[]
{\label{MHSPCenter2Edge3D}\includegraphics[height=0.2\textheight]{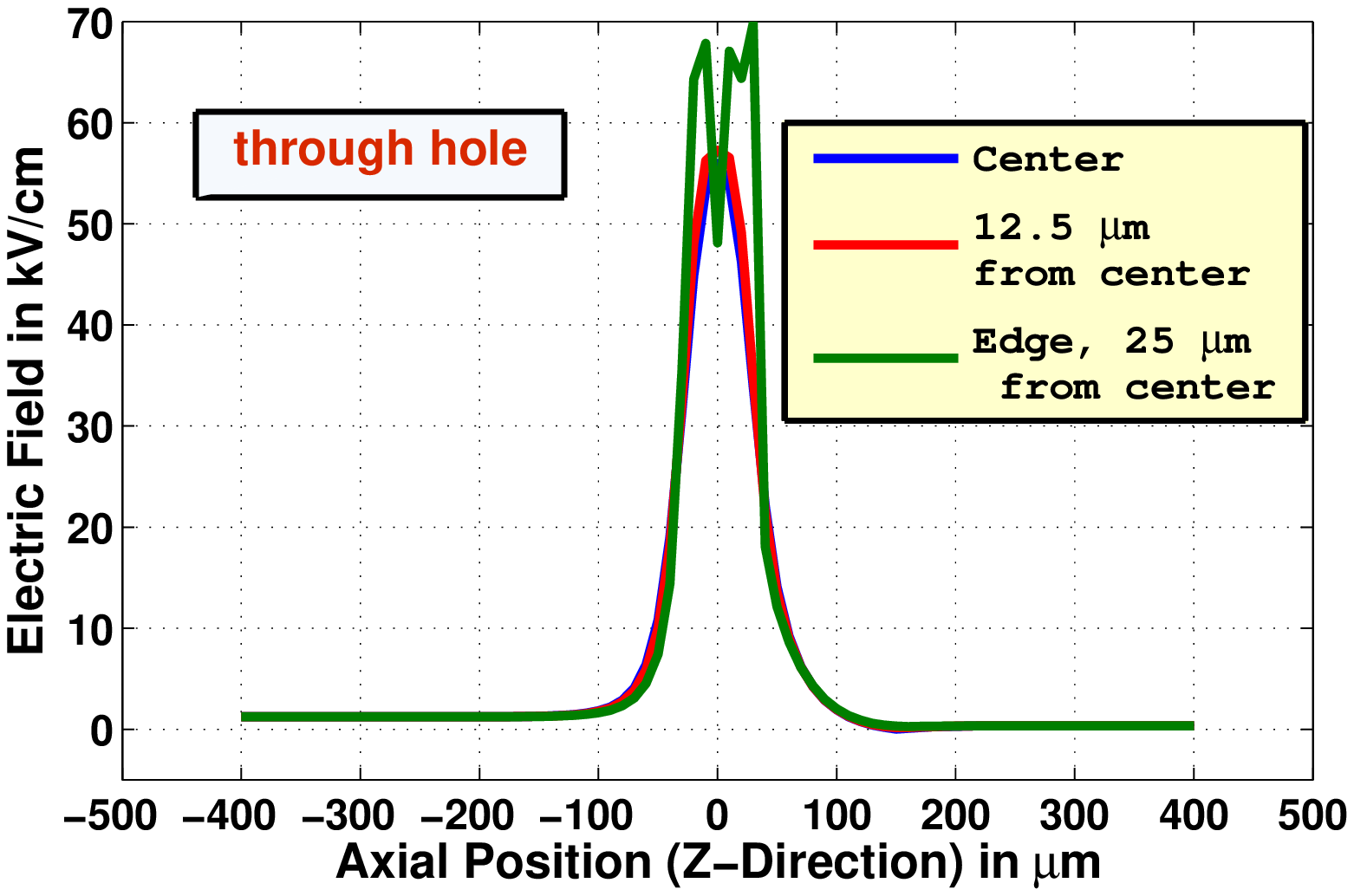}}
\caption{(a) Comparison of the total electric field of a MHSP and a GEM along the axial lines passing through the hole center for a fixed $\mathrm{V}_\mathrm{h}$ and different $\mathrm{V}_\mathrm{ac}$,
(b) Total electric field of a MHSP along an axial line passing through the hole center and two off-center lines along z-direction}
\label{FieldMHSPGEM}
\end{figure}

Figure \ref{FieldMHSPGEMCenter} presents the variation of the total electric
field along the hole axis of the MHSP for different $\mathrm{V}_\mathrm{ac}$
and a fixed $\mathrm{V}_\mathrm{h}$.
A comparison with the GEM total field reveals that, for the same
$\mathrm{V}_\mathrm{h}$, the fields for MHSP and GEM can be made identical by
assigning $\mathrm{V}_\mathrm{ac} = 0~\mathrm{V}$.
Thus it is possible to operate the MHSP in a GEM mode.
It can be also noted that an increase in $\mathrm{V}_ \mathrm{ac}$ does not
affect the maximum value of the field.

One of the main differences between a GEM and a MHSP detector is that for the
GEM, the voltage applied on the bottom induction plane is positive with respect
to the bottom grid surface, so that on the emergence from the hole, 
the electrons drift towards the induction plane where they are collected.
But in the case of a MHSP, the induction plane voltage is negatively biased
with respect to the bottom cathode strips voltage.
As a result the electrons are deflected towards the anode strips.
The field in the induction region for a GEM is higher than that for a MHSP.

Figure \ref{MHSPCenter2Edge3D} compares the field lines passing through the hole center with two off-center lines along the z-direction.
From this figure it is seen that as we proceed towards the edge of the hole, the smooth nature of the field is distorted by sharp gradients.

\begin{figure}[hbt]
\centering
\subfigure[]
{\label{FieldTopSideVac}\includegraphics[height=0.2\textheight]{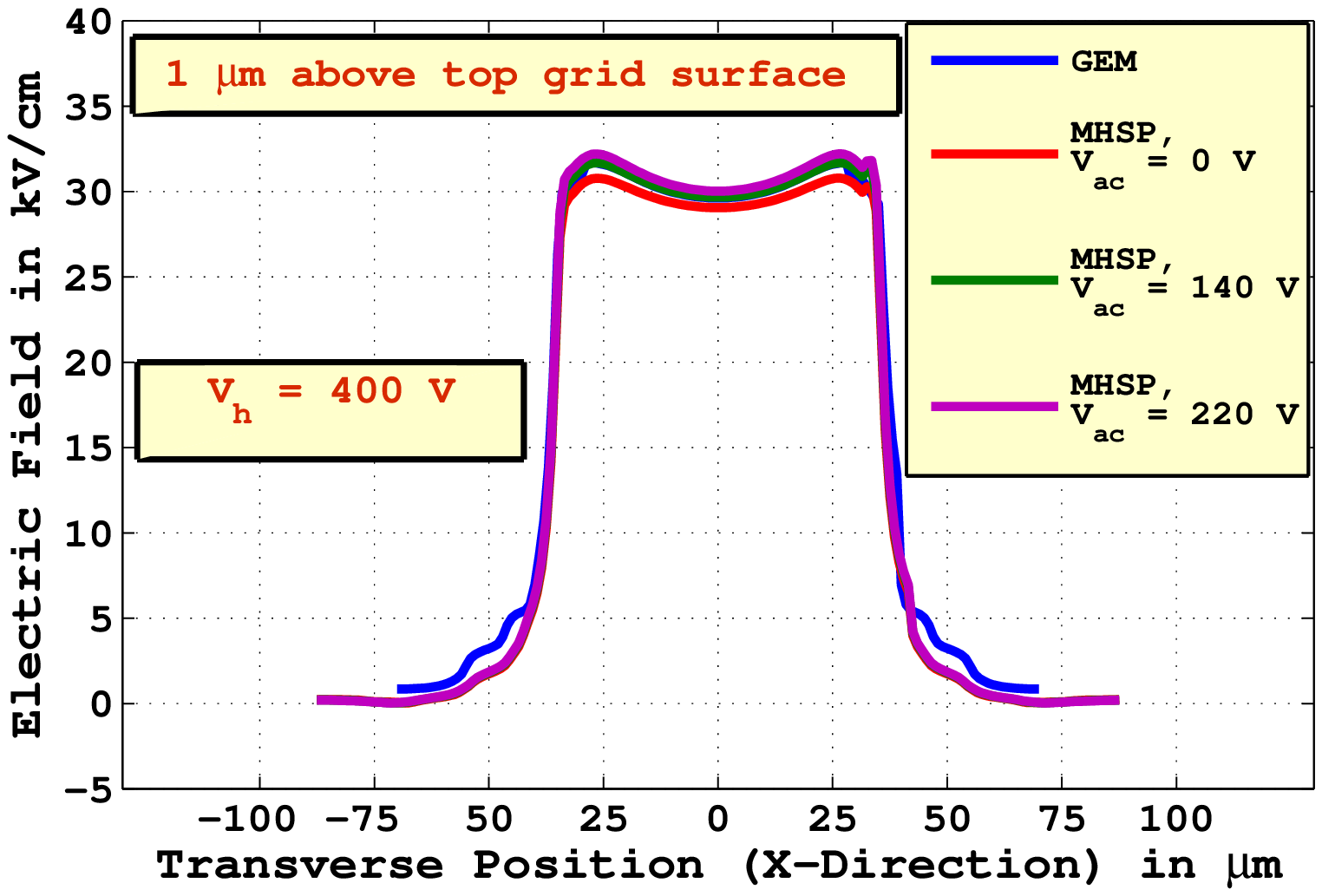}}
\subfigure[]
{\label{FieldTopSideVhole}\includegraphics[height=0.2\textheight]{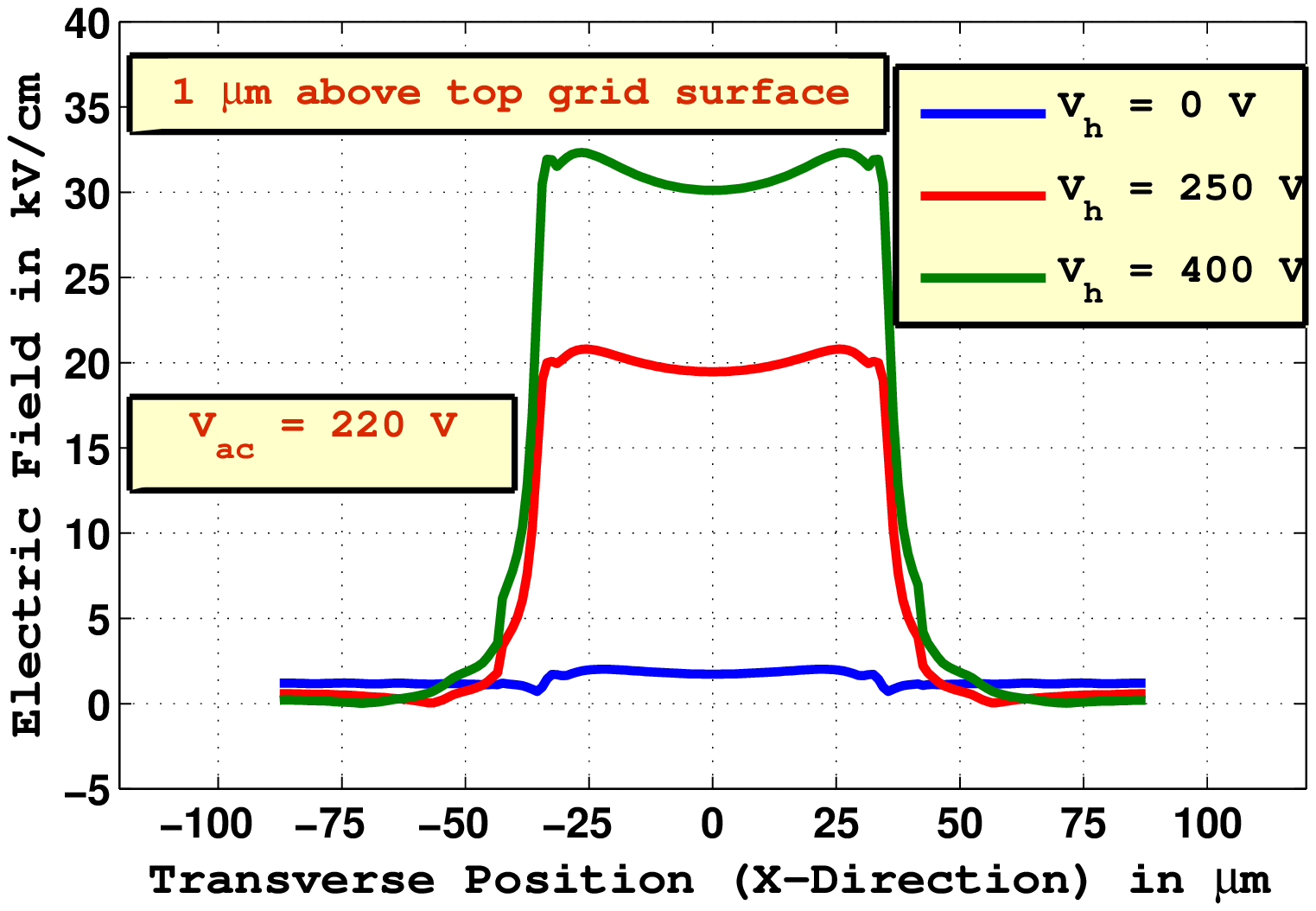}}
\caption{Variation of the electric field in the immediate proximity of the hole entrance due
to the change of (a) $\mathrm{V}_\mathrm{ac}$ for a fixed $\mathrm{V}_\mathrm{h}$, 
(b) $\mathrm{V}_\mathrm{h}$ for a fixed $\mathrm{V}_\mathrm{ac}$}
\label{FieldTopSide}
\end{figure}

Figure \ref{FieldTopSideVac} shows that the field near the grid surface (T) and in the immediate
proximity of the hole entrance is also not influenced by $\mathrm{V}_\mathrm{ac}$ and is 
same as the top surface of a GEM. If we increase the $\mathrm{V}_\mathrm{ac}$ 
from $0~\mathrm{V}$ to $220~\mathrm{V}$, the change in the value of the electric field is 
only $5 ~\%$. A similar 
observation was also made for Reverse-MHSP \cite{BreskinMHSP}. But a change in 
$\mathrm{V}_\mathrm{h}$ has a
direct impact on this field (Figure \ref{FieldTopSideVhole}). Thus the electron 
focusing into MHSP holes are not affected much by $V_\mathrm{ac}$
and can be studied as a function of $V_\mathrm{h}$ only.

\begin{figure}[hbt]
\centering
\subfigure[]
{\label{PotentialMSSide}\includegraphics[height=0.2\textheight]{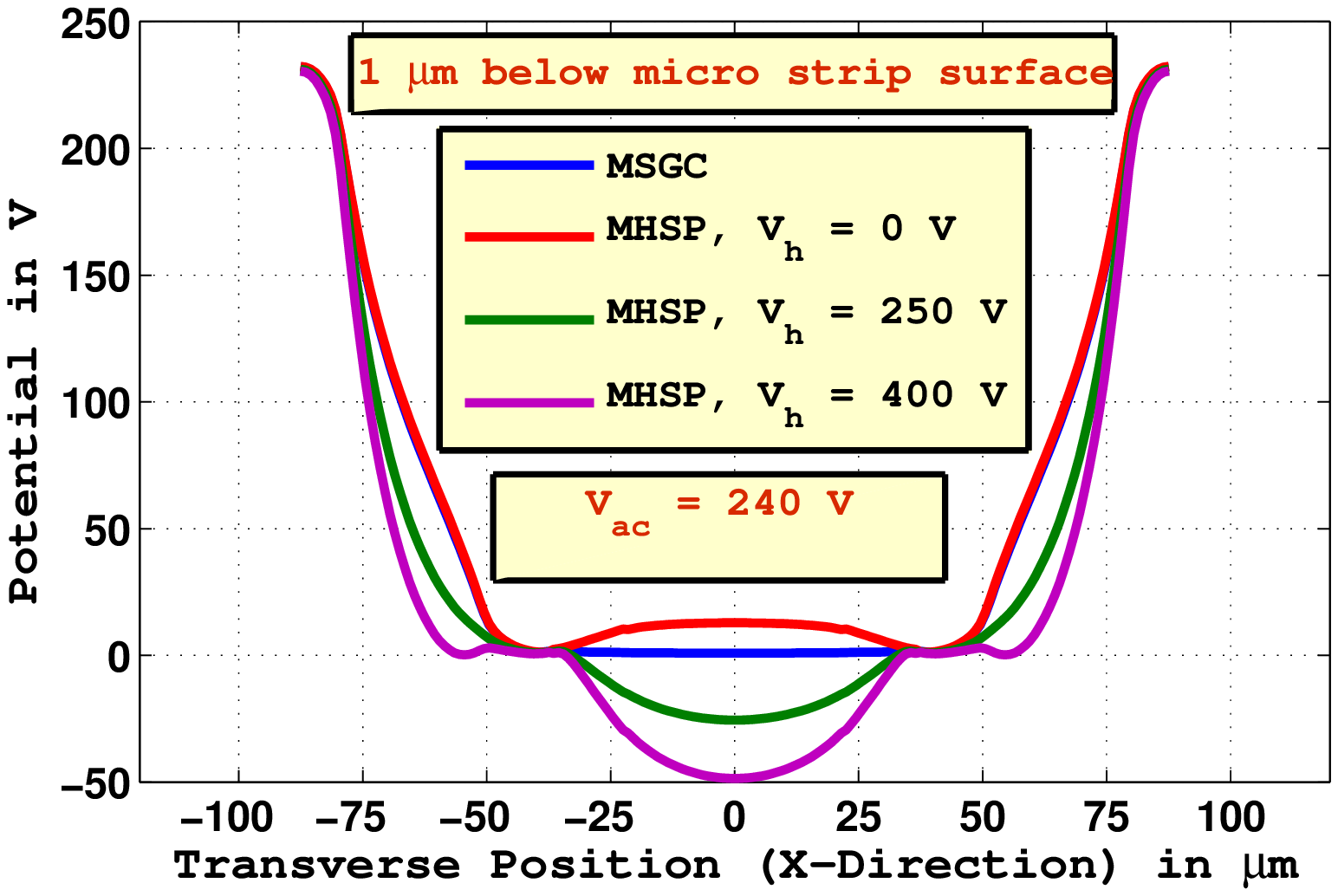}}
\subfigure[]
{\label{FieldMSSide}\includegraphics[height=0.2\textheight]{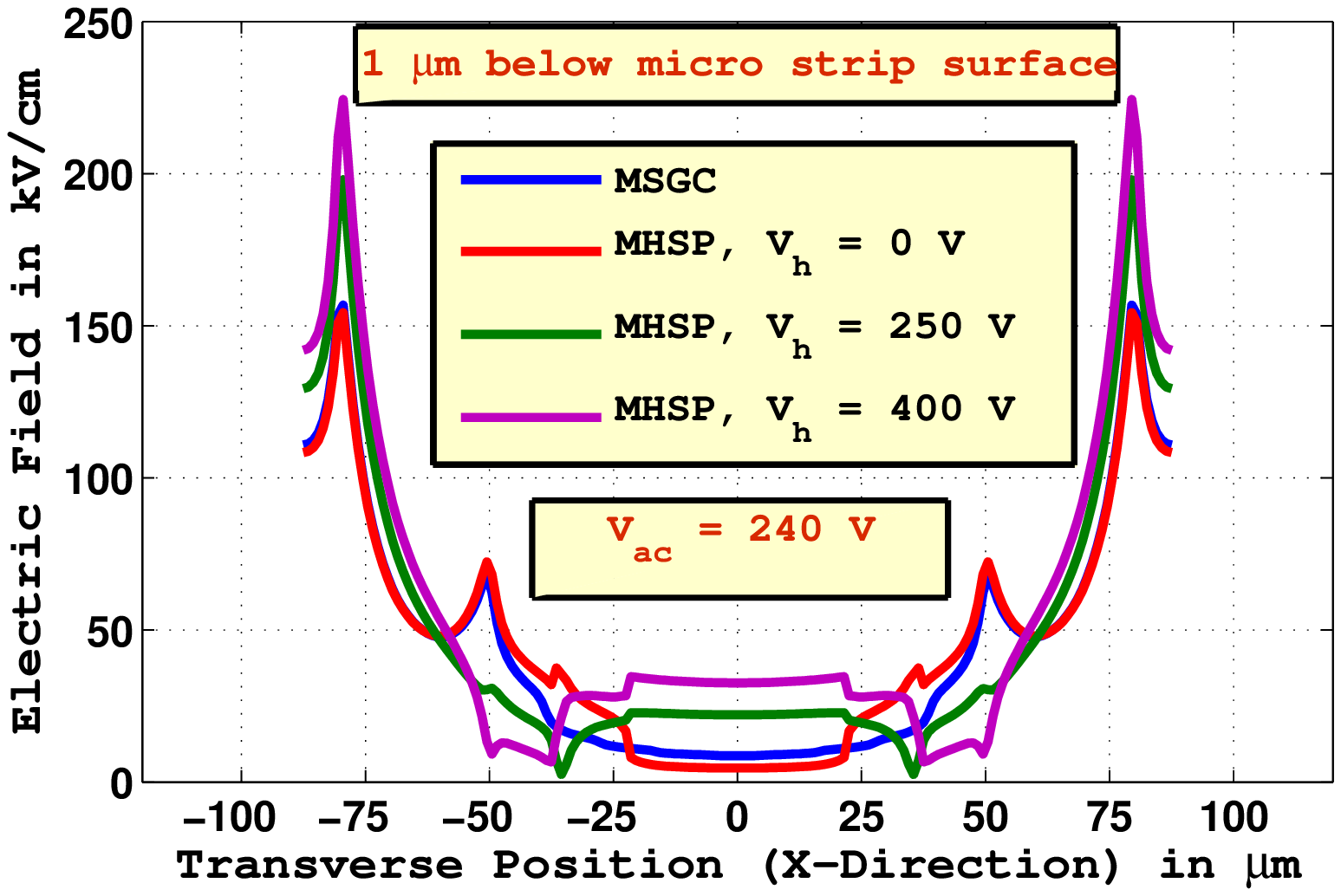}}
\caption{Effect of $\mathrm{V}_\mathrm{h}$ on the (a) Potential, (b) Electric field near the micro strip surface, for a fixed $\mathrm{V}_\mathrm{ac}$}
\label{MSSide}
\end{figure}

In figure \ref{MSSide} we have depicted the variation of the potential and the electric field on the
Micro Strip surface of a MHSP due to variation in $\mathrm{V}_\mathrm{h}$ for a 
fixed $\mathrm{V}_\mathrm{ac}$. 
Without any hole voltage, the bottom microstrip surface of the MHSP, acts as an ordinary
MSGC. It is further observed that potential and field are both strongly affected by a 
variation in $\mathrm{V}_\mathrm{h}$. As a result, the multiplication factor in 
second amplification
stage ($\mathrm{g}_\mathrm{s}$) and the collection efficiency of the anode 
not only depends on $\mathrm{V}_\mathrm{ac}$, but also on $\mathrm{V}_\mathrm{h}$. 

\subsection{Electron Collection Efficiency}
\label{sec:efficiency}

\begin{figure}[hbt]
\centering
\subfigure[]
{\label{DriftLines1m}\includegraphics[height=0.2\textheight]{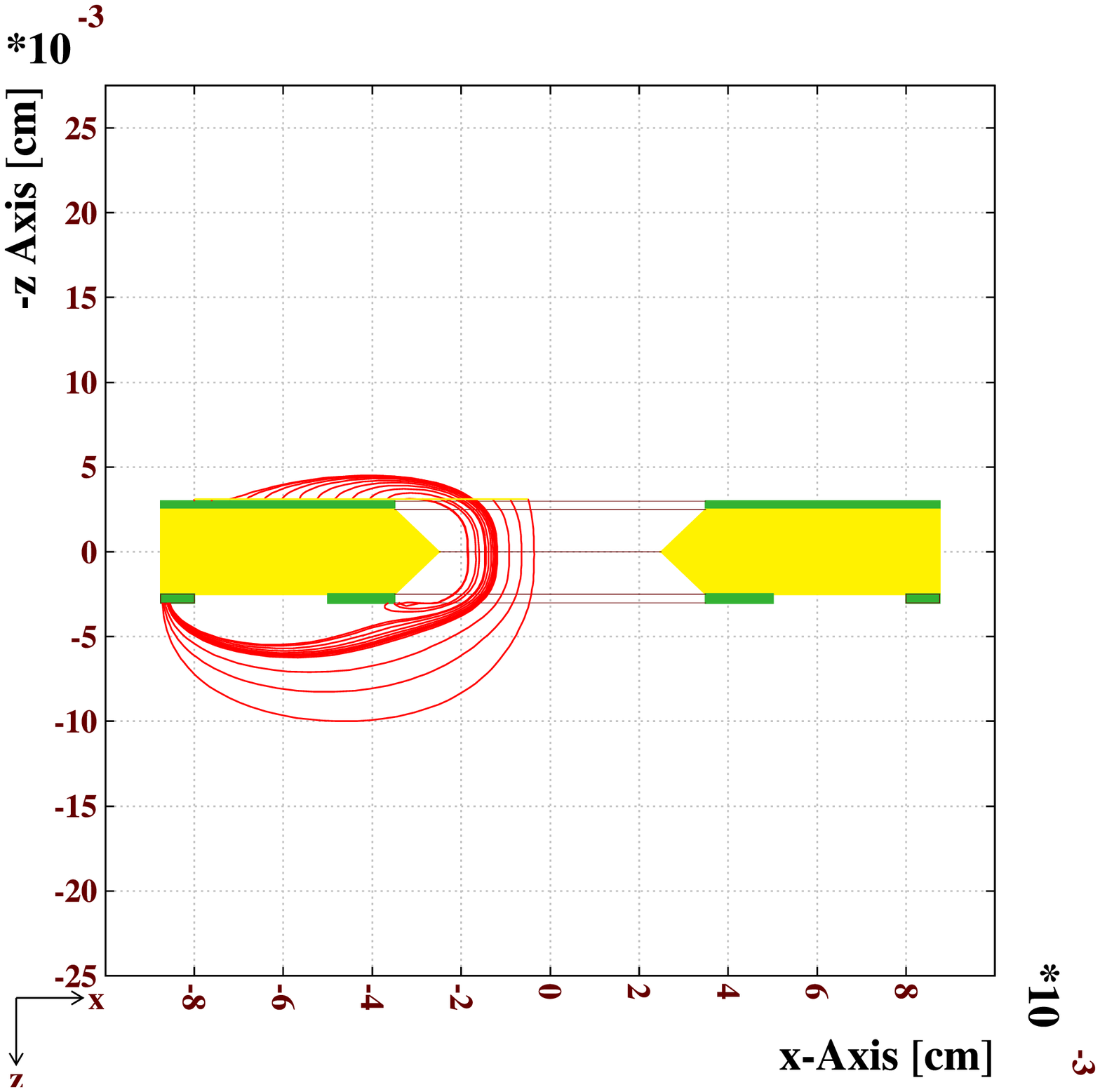}}
\subfigure[]
{\label{DriftLines240m}\includegraphics[height=0.2\textheight]{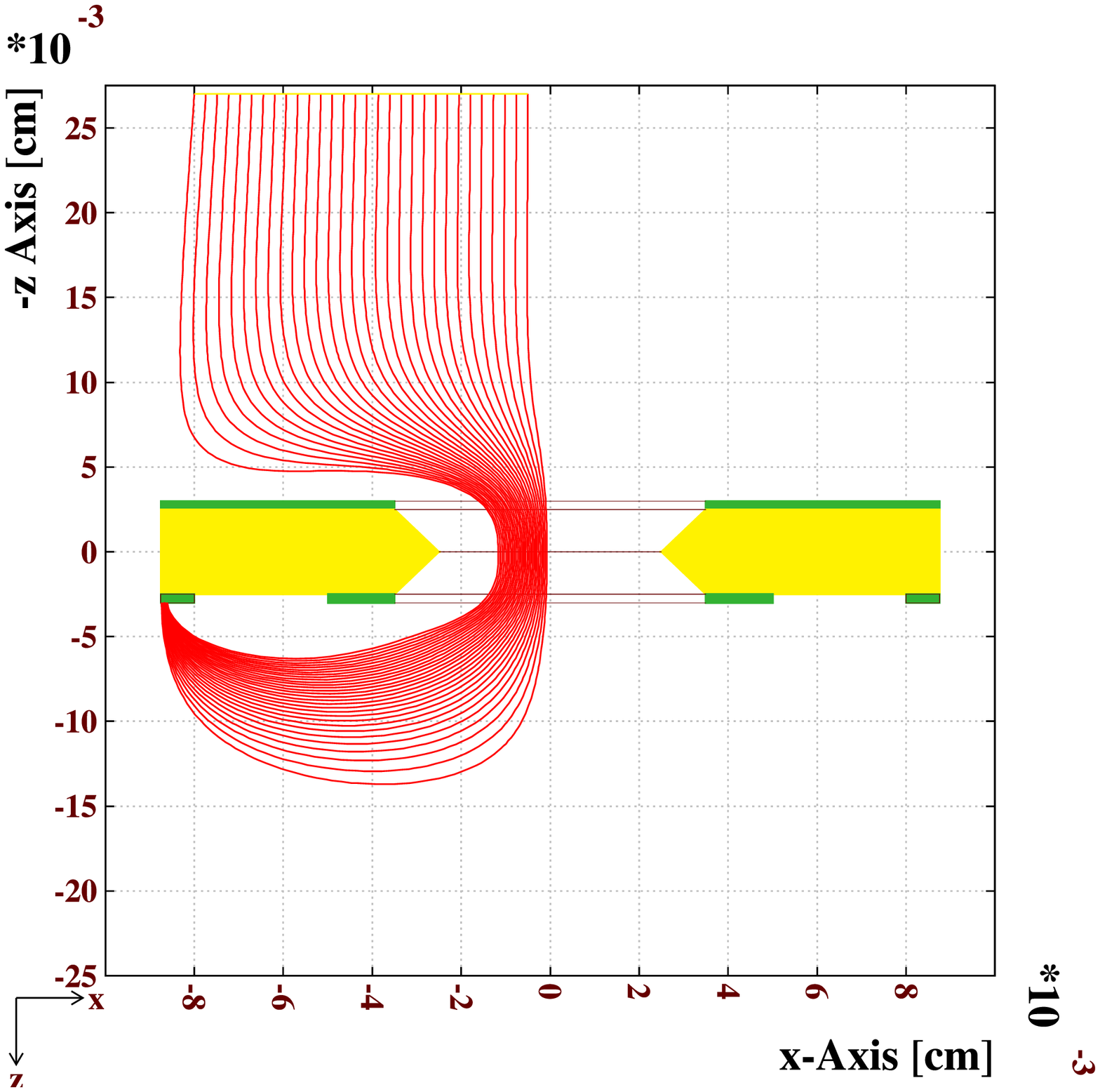}}
\caption{Electron drift lines using RKF method from pre-defined track; (a) $1~\mu\mathrm{m}$ and (b) $240~\mu\mathrm{m}$ above the top grid surface}
\label{DriftLines}
\end{figure}

An $\mathrm{Ar}-\mathrm{CO_2}$ gas mixture ($70:30$) at $293~\mathrm{K}$
and $1 ~\mathrm{atmosphere}$ has been considered here.
We have defined two tracks of primary-ionization electrons in the drift region: 
1) a set of 30 electrons starting at $240~\mu\mathrm{m}$ above the MHSP top grid and 
2) a set of 20 electrons starting at a position $1~\mu\mathrm{m}$ above the grid. 
In the following discussions the drift lines are estimated
using simple Runge-Kutta-Fehlberg (RKF) method and thus the electron diffusion in gas is 
ignored (figure \ref{DriftLines}).

\begin{figure}[hbt]
\centering
\includegraphics[height=.2\textheight]{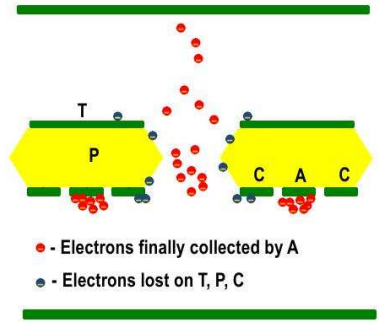}
\caption{{\label{schematic2}}The end point of electrons on different electrodes for MHSP}
\end{figure}

\begin{figure}[hbt]
\centering
\subfigure[]
{\label{MHSP3DEfficiency-1a}\includegraphics[height=0.2\textheight]{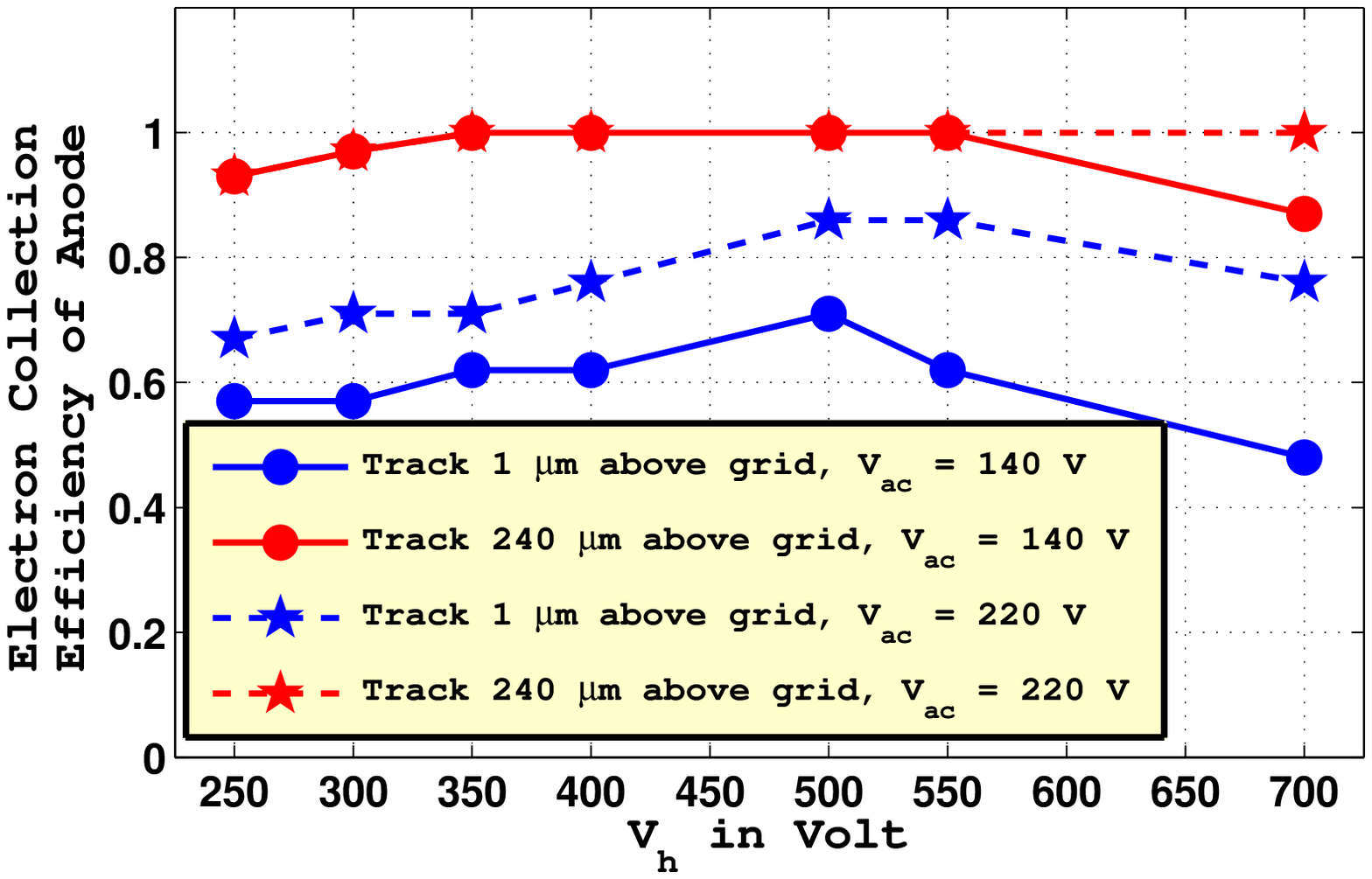}}
\subfigure[]
{\label{MHSP3DEfficiency-1b}\includegraphics[height=0.2\textheight]{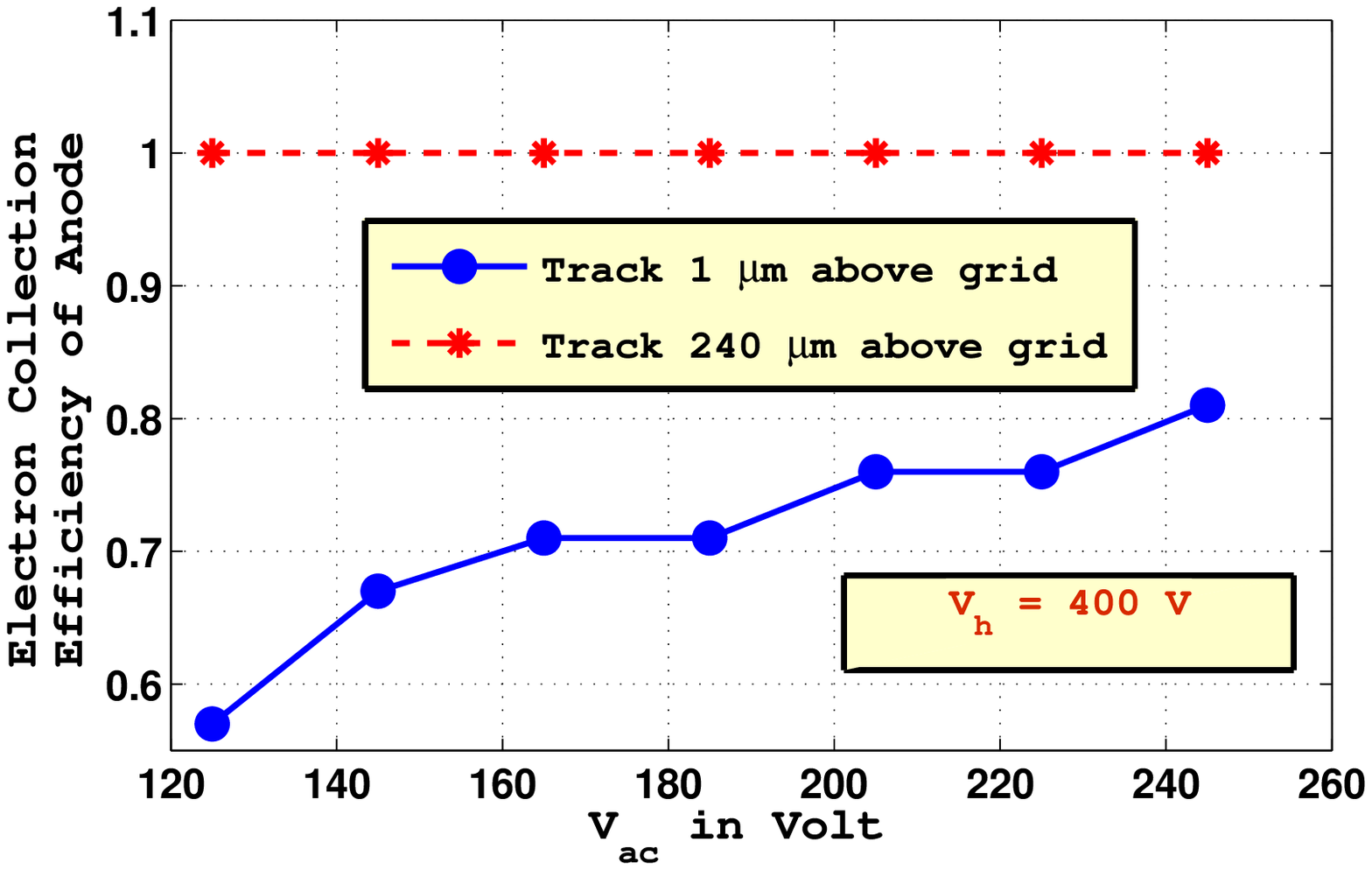}}
\caption{Variation of electron collection efficiency of anode, 
(a) with $\mathrm{V}_\mathrm{h}$ (b) with $\mathrm{V}_\mathrm{ac}$}
\label{MHSP3DEfficiency-1}
\end{figure}

\begin{figure}[hbt]
\centering
\subfigure[]
{\label{MHSP3DEfficiency-2a}\includegraphics[height=0.2\textheight]{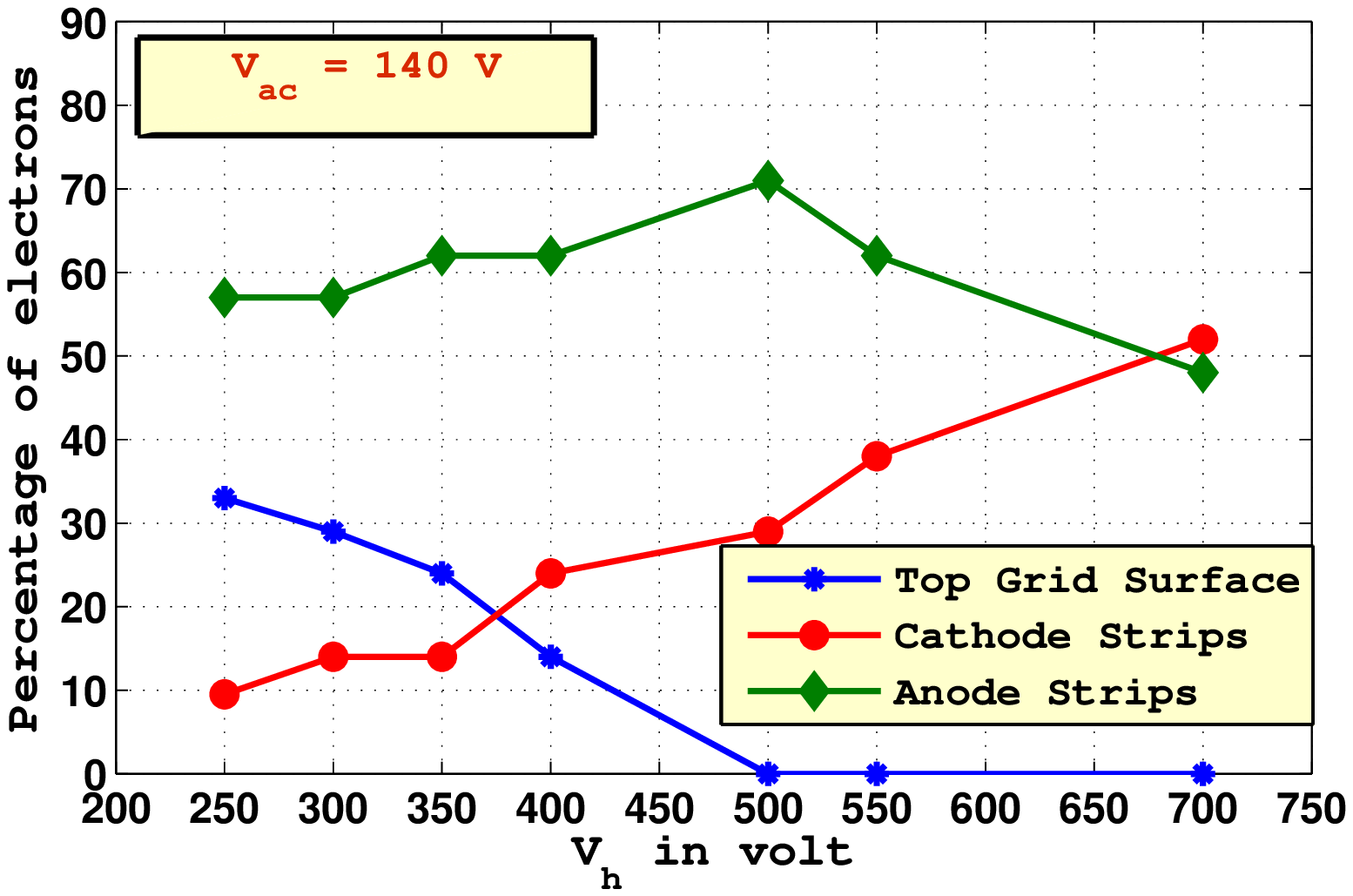}}
\subfigure[] 
{\label{MHSP3DEfficiency-2b}\includegraphics[height=0.2\textheight]{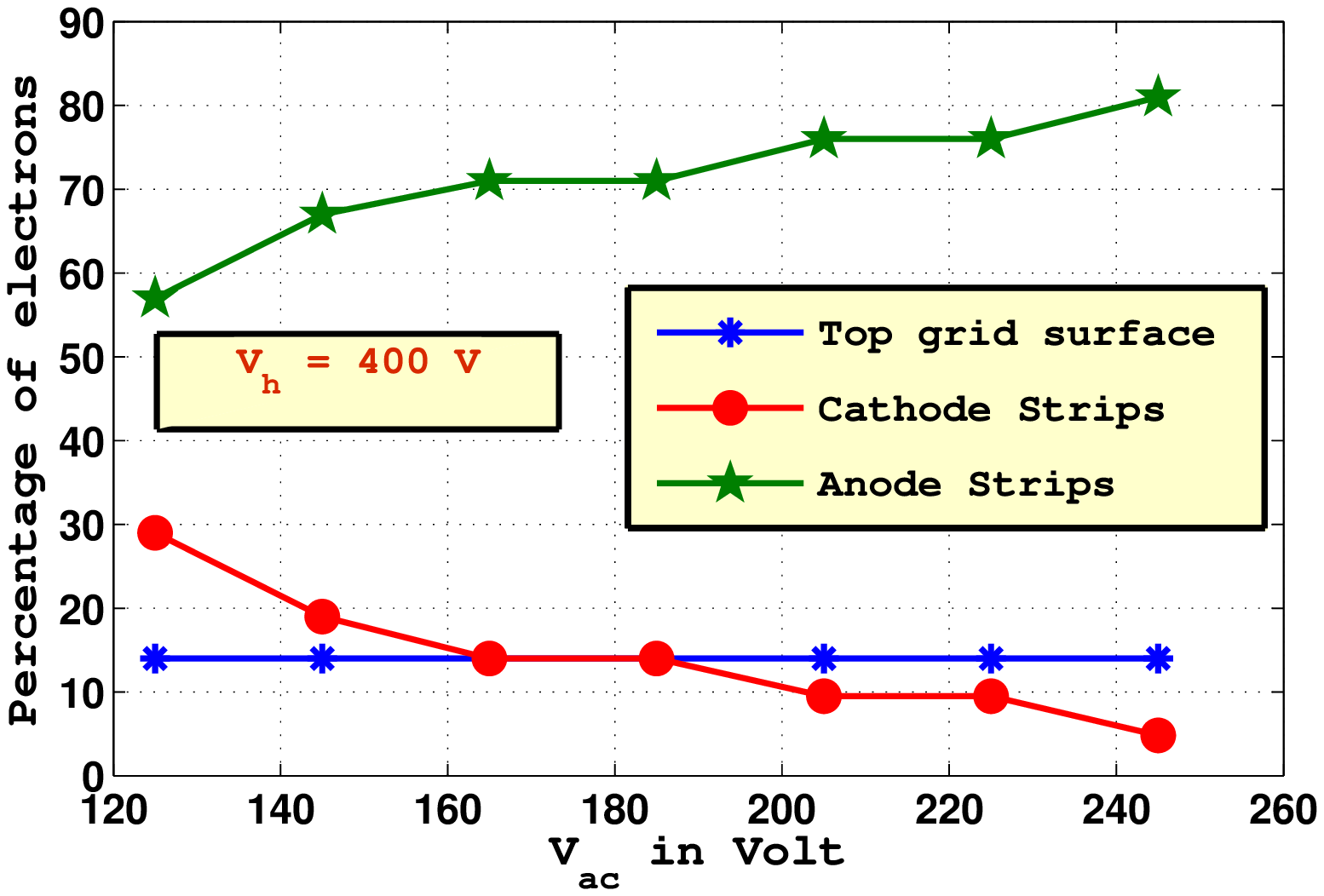}}
\caption{Percentage of electrons collected on different electrodes,  
(a) dependence on $\mathrm{V}_\mathrm{h}$ for a fixed $\mathrm{V}_\mathrm{ac}$, 
(b) dependence on $\mathrm{V}_\mathrm{ac}$ for a fixed $\mathrm{V}_\mathrm{h}$}
\label{MHSP3DEfficiency-2}
\end{figure}

Depending on the voltage settings of different electrodes, some electrons are lost on 
the top grid surface, dielectric and the cathode strips as shown schematically
in figure \ref{schematic2}. The variation of the electron collection efficiency of 
the anode 
with $\mathrm{V}_\mathrm{h}$ for a fixed $V_\mathrm{ac}$ and with
$\mathrm{V}_\mathrm{ac}$ for a fixed $\mathrm{V}_\mathrm{h}$ are presented in 
figure \ref{MHSP3DEfficiency-1a} and figure \ref{MHSP3DEfficiency-1b} respectively. 
Variations of $\mathrm{V}_\mathrm{drift}$ and $\mathrm{V}_\mathrm{ind}$ 
is also expected to affect the parameter, but have not been considered in the present study. 
For a fixed $\mathrm{V}_\mathrm{ac}$, a minimum $\mathrm{V}_\mathrm{h}$  is required to focus
the electrons towards the hole, otherwise the electrons are lost on the top of
the grid surface. Beyond a certain $\mathrm{V}_\mathrm{h}$  all the electrons are focused
towards the hole and finally collected by the anode strips. But if we increase
the $\mathrm{V}_\mathrm{h}$  even more, some of the electrons which start their journey close to
the grid surface (mainly the electrons whose drift paths are near the edge of the
hole) end their journey at the cathode strips (figure \ref{MHSP3DEfficiency-2a}) since for them 
the anode strips voltage is not sufficient to pull them. As a result, the efficiency drops
(blue lines in figure \ref{MHSP3DEfficiency-1a}). We can increase this efficiency once again 
by increasing $\mathrm{V}_\mathrm{ac}$ (blue line in figure \ref{MHSP3DEfficiency-1b}, since for a 
particular $\mathrm{V}_\mathrm{h}$, increase of the $\mathrm{V}_\mathrm{ac}$ attracts 
the electrons more towards the anode strip (figure \ref{MHSP3DEfficiency-2b}).

The electrons which start their journey in the middle of the drift region,
mainly drift through the central part of the hole. So most of these electrons
reach the anode strips safely (red lines of Figure \ref{MHSP3DEfficiency-1a} and 
figure \ref{MHSP3DEfficiency-1b}).

From the above graphs, it is seen that the electron collection
efficiency can be maximized with proper optimization of $\mathrm{V}_\mathrm{h}$ and 
$\mathrm{V}_\mathrm{ac}$. For this particular geometry, a $\mathrm{V}_\mathrm{h}$ 
of $500~\mathrm{V}$ is suitable
(collection efficiency $80~\%$) for two sets of $\mathrm{V}_\mathrm{ac}$ (140 V and 220 V), 
studied here. 
At this value of $\mathrm{V}_\mathrm{h}$, an increase of $\mathrm{V}_\mathrm{ac}$ 
certainly improves the efficiency,
but this choice of voltage is likely to be governed by the sparking limit.

\subsection{Gain}
\label{sec:gain}

The effective gain of electrons for a particular track is obtained as 
\begin{eqnarray}
\mathrm{g}_\mathrm{eff} = \epsilon_\mathrm{prim} \times \mathrm{g}_\mathrm{mult} \times \epsilon_\mathrm{sec} 
\end{eqnarray}

\noindent where $\epsilon_\mathrm{prim}$ is the primary electron collection efficiency 
and is the probability for a primary
electron to reach the hole region. $\mathrm{g}_\mathrm{mult}$ is the multiplication factor of the
electrons throughout their trajectories. For a GEM, the multiplication occurs 
only inside the hole ($\mathrm{g}_\mathrm{h}$ in figure \ref{schematic} and thus, 
$\mathrm{g}_\mathrm{mult} = \mathrm{g}_\mathrm{h}$). 
In case of MSGC, $\mathrm{V}_\mathrm{ac}$ is responsible for the electron multiplication
near the micro strip surface ($\mathrm{g}_\mathrm{mult} = \mathrm{g}_\mathrm{s}$). As designed, 
the MHSP combines these two stages of multiplication
(figure \ref{schematic}). So in this case, 
$\mathrm{g}_\mathrm{mult} = \mathrm{g}_\mathrm{h} \times \mathrm{g}_\mathrm{s}$. $\epsilon_\mathrm{sec}$ 
is the secondary electron collection
efficiency of the readout electrode. It may be noted here that we have considered only
the collected charge at the anode strips (A) to determine effective gain .

The discussion in the previous section suggests that electrons arising from different 
positions of the drift region behave quite differently. Since
in an experiment, the gamma rays from a radiation source can liberate 
the primary electrons in
different parts of the drift region, we choose four tracks at different distances
above the top grid surface ($1~\mu\mathrm{m}$, $10~\mu\mathrm{m}$, $500\mu\mathrm{m}$ 
and $1~\mathrm{mm}$). The total gain
($g_\mathrm{t}$) is the average effective gain of the electrons from these four tracks.

The gas mixture considered in this work is a
Penning mixture. After considering results using two transfer
rates, $56\%$ (extrapolated value from \cite{Penning}) and $70\%$ (a guess work),
we chose to carry out the rest of the calculations with the higher value since
it agreed well with the experimental data. This issue, however, needs further investigation.

\begin{figure}[hbt]
\centering
\subfigure[]
{\label{GEMGain3D}\includegraphics[height=0.2\textheight]{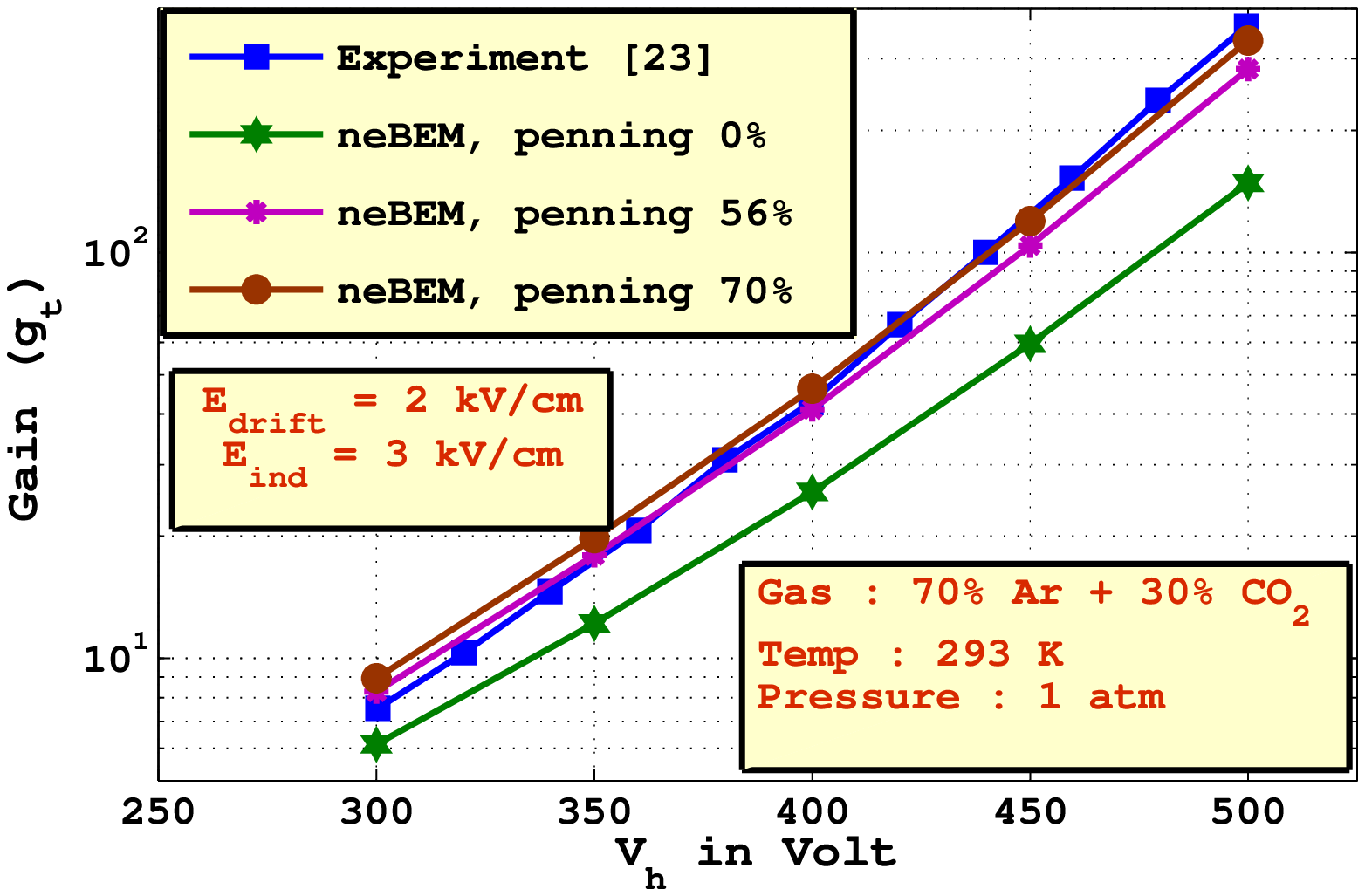}}
\subfigure[]
{\label{MHSPGain3DVhole-2}\includegraphics[height=0.2\textheight]{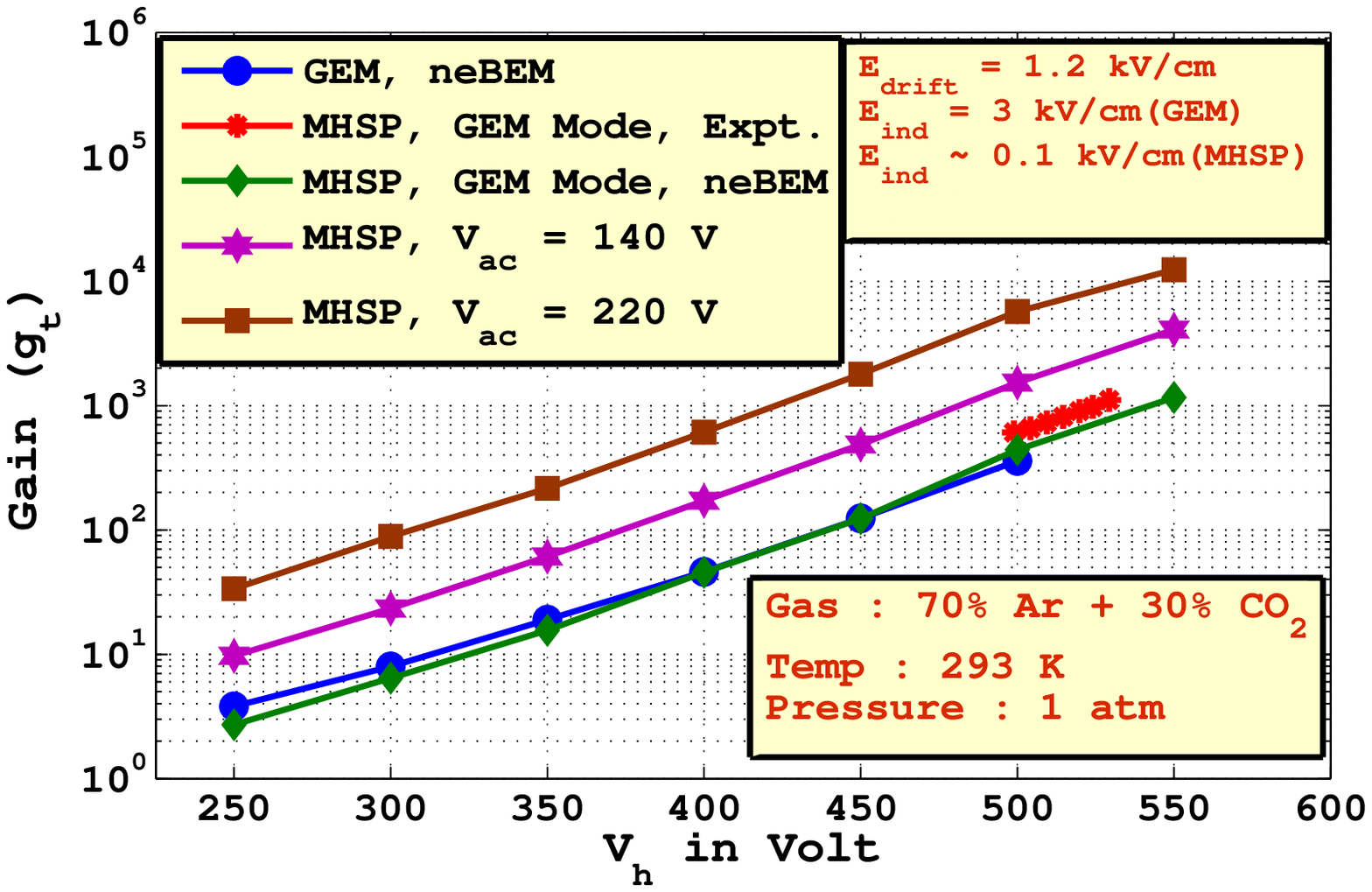}}
\caption{Variation of $\mathrm{g}_\mathrm{t}$ with $\mathrm{V}_\mathrm{h}$ under different conditions for (a) GEM , (b) MHSP}
\label{MHSPGain3DVhole}
\end{figure}

The variation of the gain ($\mathrm{g}_t$) with $\mathrm{V}_\mathrm{h}$ for a GEM, is shown in figure \ref{GEMGain3D}.
For a MHSP, the same variation of 
$\mathrm{g}_\mathrm{t}$ 
is depicted in figure \ref{MHSPGain3DVhole-2}. When the MHSP is operated in a GEM mode 
($\mathrm{V}_\mathrm{ac}~=~0~\mathrm{V}$, $\mathrm{g}_\mathrm{s}~=~1$ and 
thus $\mathrm{g}_\mathrm{mult}~=~\mathrm{g}_\mathrm{h}$), 
$\mathrm{g}_\mathrm{t}$ is 
similar to that obtained with a single GEM. But for the same variation 
of $\mathrm{V}_\mathrm{h}$, $\mathrm{g}_\mathrm{t}$ increases with
the increase of $\mathrm{V}_\mathrm{ac}$. The trend obtained from the present 
simulated estimates is
similar to that observed in experiment by Veloso et al. \cite{VelosoMHSP3}.

\begin{figure}[hbt]
\centering
\subfigure[]
{\label{MHSPGain3DVac-1}\includegraphics[height=0.2\textheight]{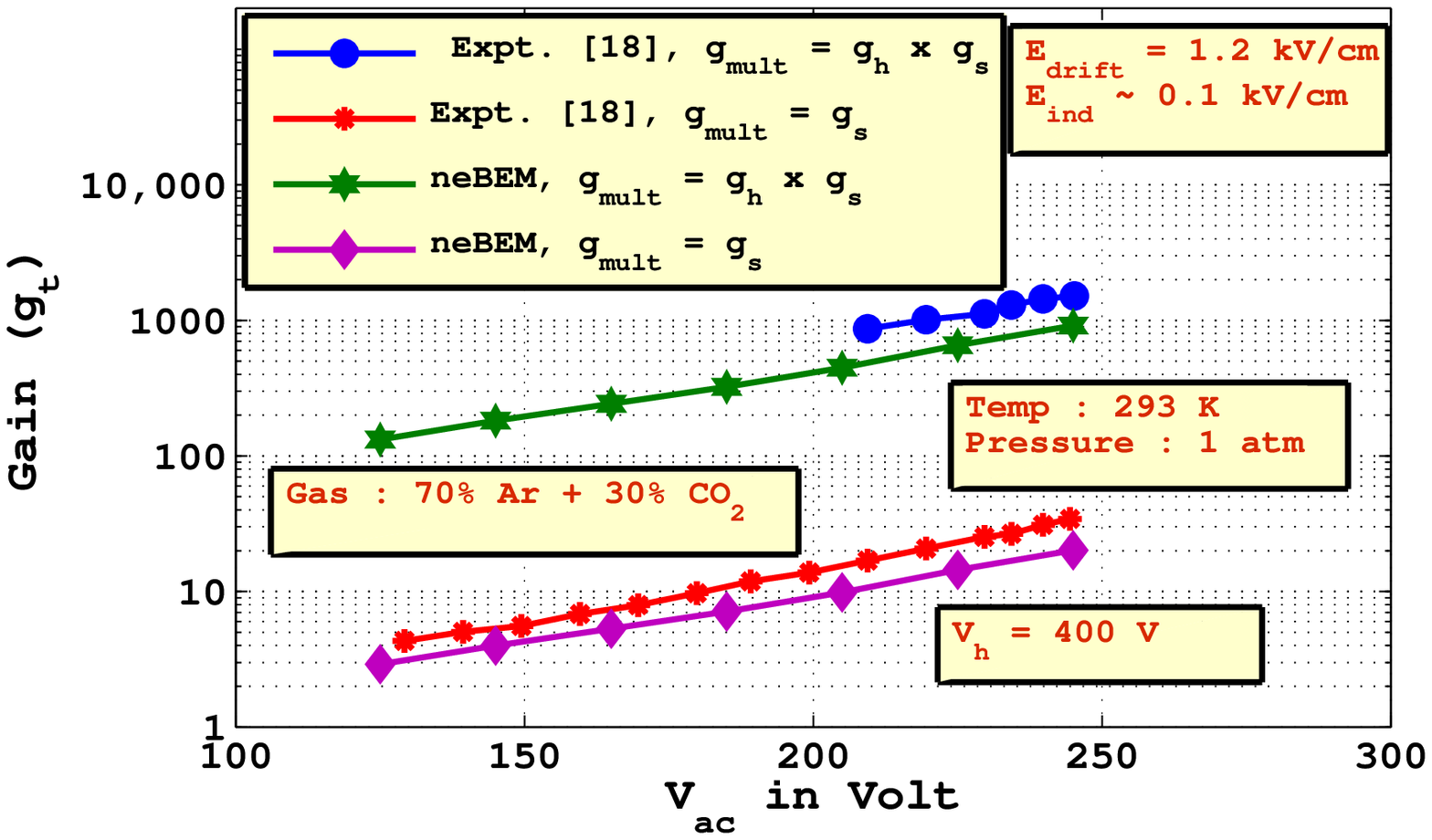}}
\subfigure[]
{\label{MHSPGain3DVac-2}\includegraphics[height=0.2\textheight]{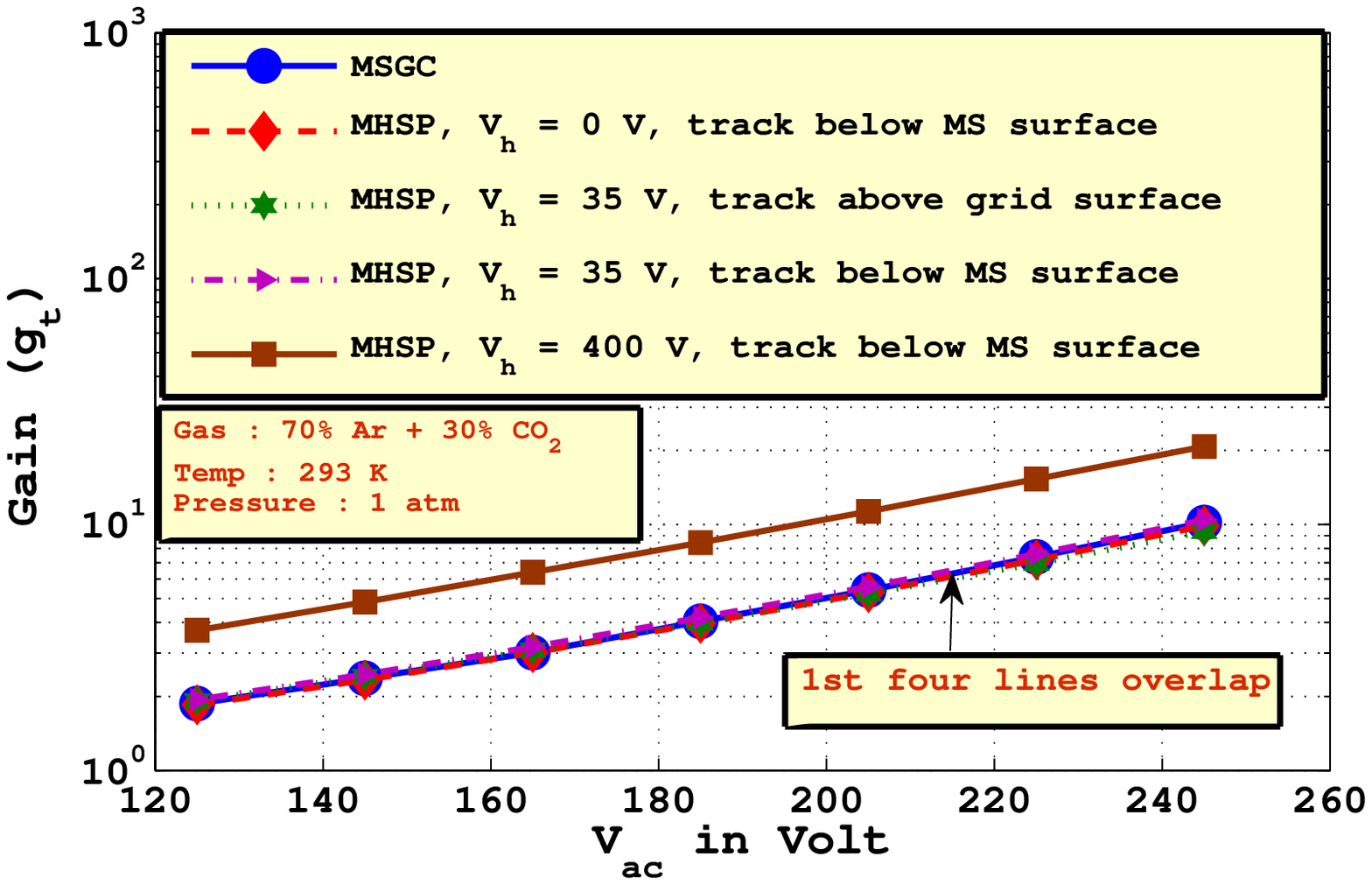}}
\caption{Variation of $\mathrm{g}_\mathrm{t}$ with $\mathrm{V}_\mathrm{ac}$ under different conditions}
\label{MHSPGain3DVac}
\end{figure}

In Figure \ref{MHSPGain3DVac-1}, we present the variation in $\mathrm{g}_\mathrm{t}$ due 
to variation in $\mathrm{V}_\mathrm{ac}$ for a fixed $\mathrm{V}_\mathrm{h}$. Since, the gain for this 
fixed $\mathrm{V}_\mathrm{h}$ can be calculated from the above procedure (MHSP 
in GEM mode) 
we can make a rough estimate of the gain ($\mathrm{g}_\mathrm{mult} = \mathrm{g}_\mathrm{s}$) 
in the 2nd amplification stages only. 
The experimental results \cite{VelosoMHSP2} verify this nature.

Experimentally it was considered in the work of Veloso et al. \cite{VelosoMHSP1} that a fraction of the incident X-rays 
can interact in the induction region, below the MHSP. The primary electron 
clouds from these events experience only one stage of charge 
multiplication at the micro strip anodes ($\mathrm{g}_\mathrm{mult}~=~\mathrm{g}_\mathrm{s}, 
\mathrm{g}_\mathrm{h}~=~1$). 
When $\mathrm{V}_\mathrm{h}~=~0~\mathrm{V}$, no electrons from the drift region can reach 
the micro strip surface and in this case, $\mathrm{g}_\mathrm{t}$ of a MHSP 
is of the same order of a single MSGC, as 
shown in the Figure \ref{MHSPGain3DVac-2}. When $\mathrm{V}_\mathrm{h}$ is adequate
for only electron transmission, but not for hole multiplication 
(for example, $\mathrm{V}_\mathrm{h}~=~35~\mathrm{V}$), 
then also the amplification factor of electrons from above or below the MHSP
depend only on $\mathrm{V}_\mathrm{ac}$.  
The gain ($\mathrm{g}_\mathrm{t}$) of the above four graphs in figure \ref{MHSPGain3DVac-2} 
is of the same order. But $\mathrm{V}_\mathrm{h}$ beyond a certain value has an          
effect not only on the electric field inside the hole but also on the micro strip surface 
(figure \ref{MSSide}) 
which affects $\mathrm{g}_\mathrm{s}$ and thus $\mathrm{g}_\mathrm{t}$. 
In this situation ($\mathrm{V}_\mathrm{h}~=~400~\mathrm{V}$ in figure \ref{MHSPGain3DVac-2}), 
$\mathrm{g}_\mathrm{t}$ is higher than that of above four graphs, but of the same order
as obtained in figure \ref{MHSPGain3DVac-1} (red and purple line). 

From the above gain study, it is seen that the total gain of a MHSP depends on the potential
difference of two different amplification stages. With proper optimization of 
$\mathrm{V}_\mathrm{h}$ and $\mathrm{V}_\mathrm{ac}$, the total gain of MHSP
can be made higher than that of a single GEM or a MSGC.

\subsection{Comparison between RKF results and MC results}
\label{RKF-MC}

\begin{figure}[hbt]
\centering
\subfigure[]
{\label{Avalanche}\includegraphics[height=0.2\textheight]{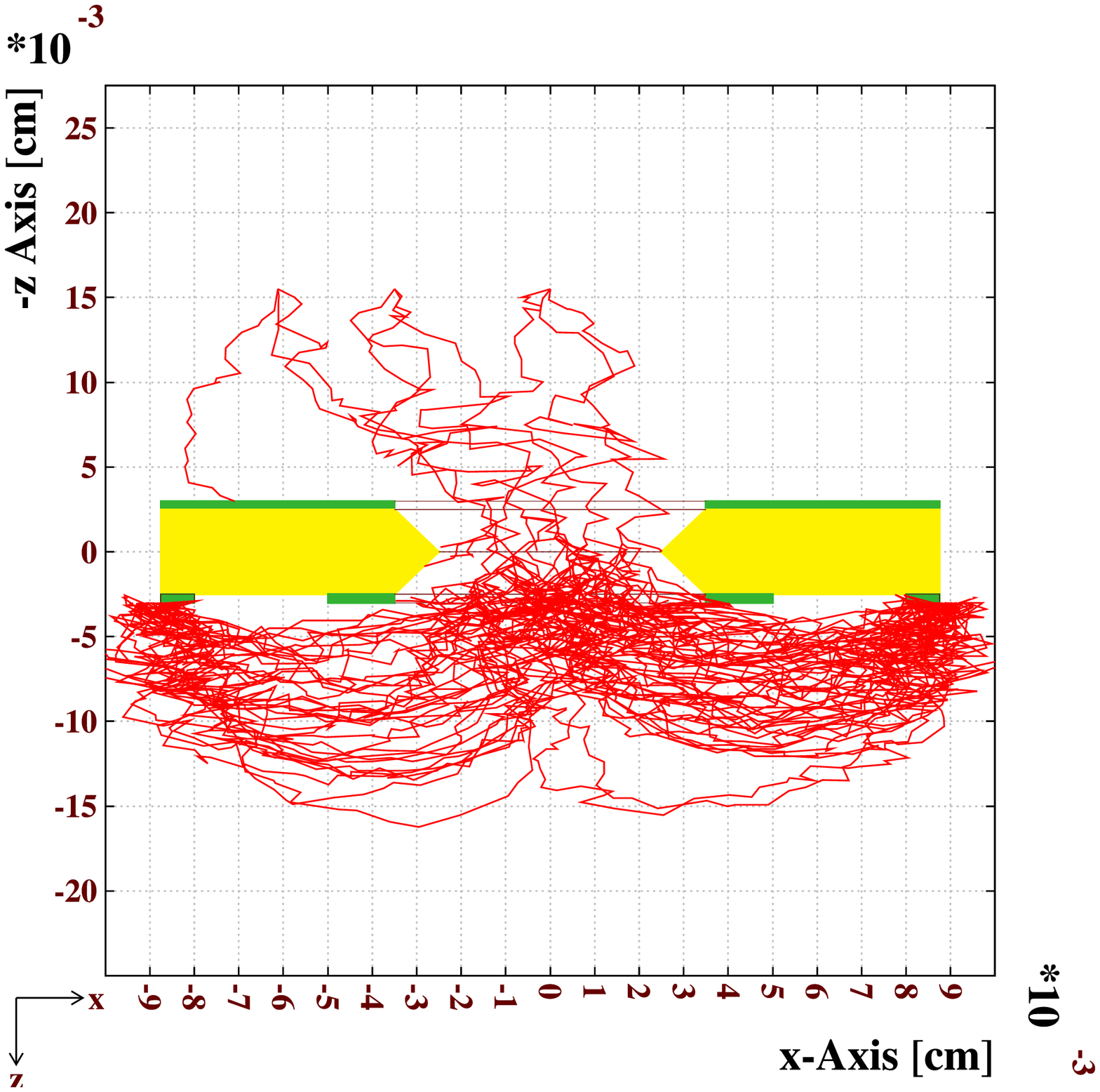}}
\subfigure[]
{\label{primary}\includegraphics[height=0.2\textheight]{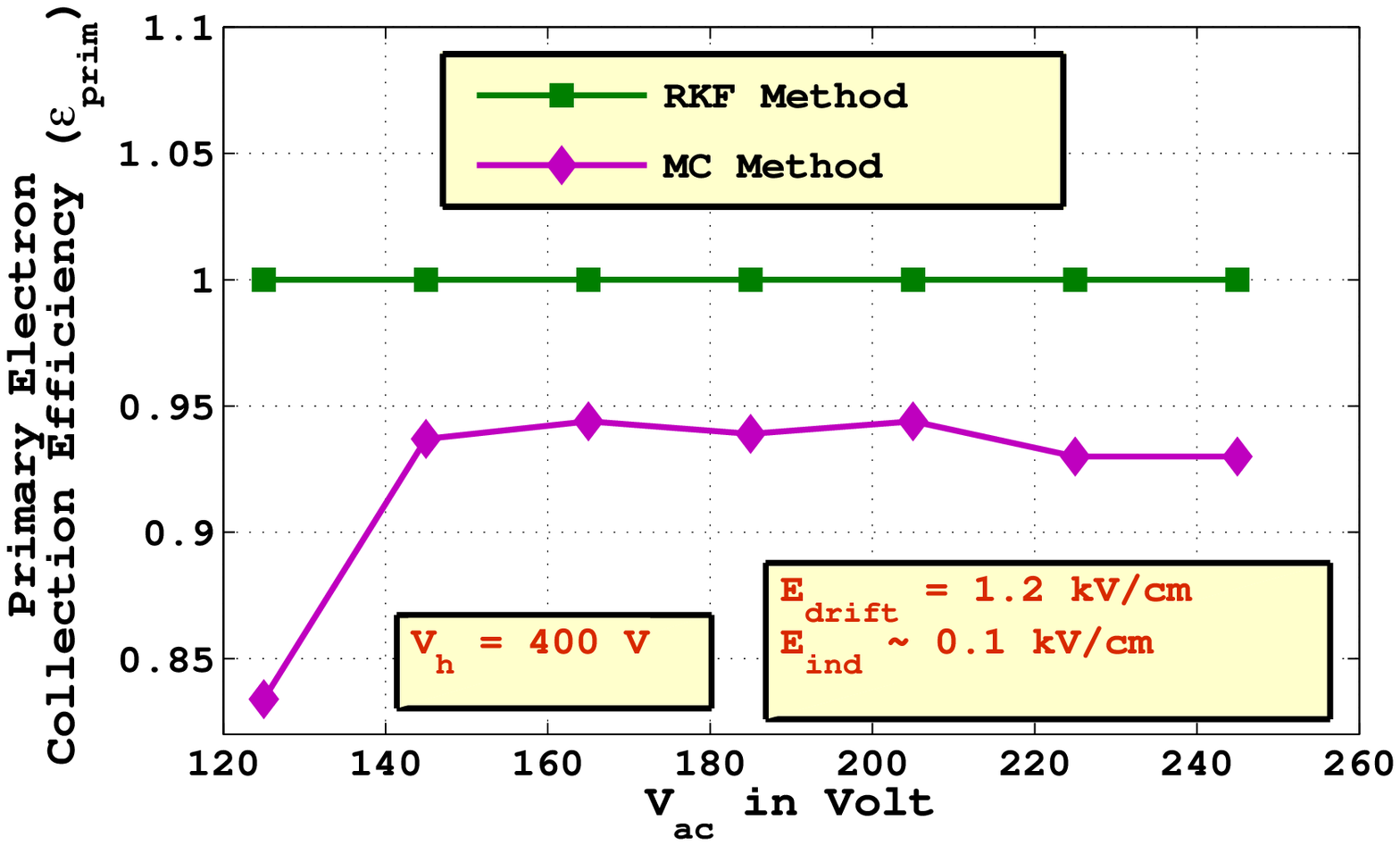}}\\
\subfigure[]
{\label{secondary}\includegraphics[height=0.2\textheight]{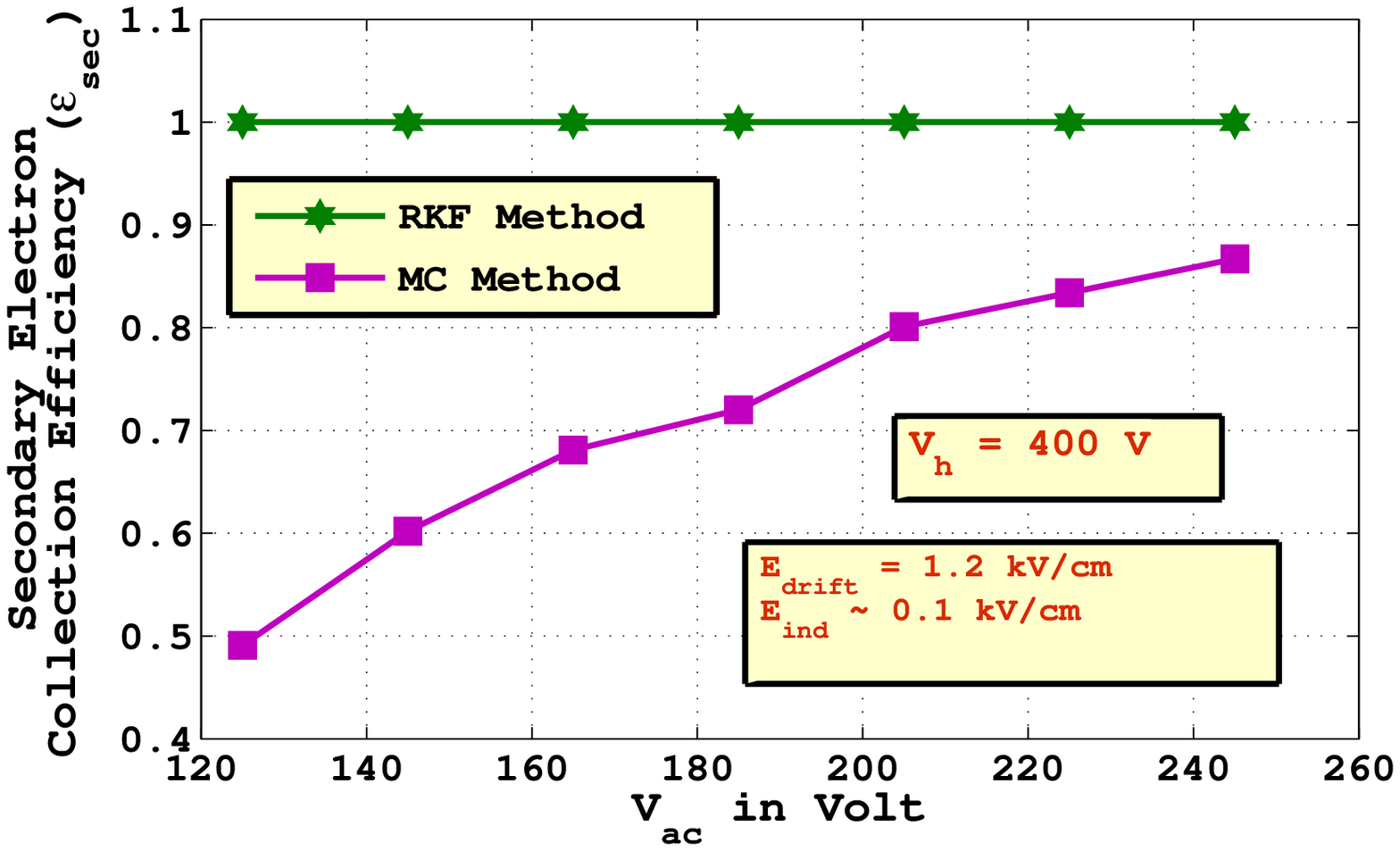}}
\subfigure[]
{\label{gain}\includegraphics[height=0.2\textheight]{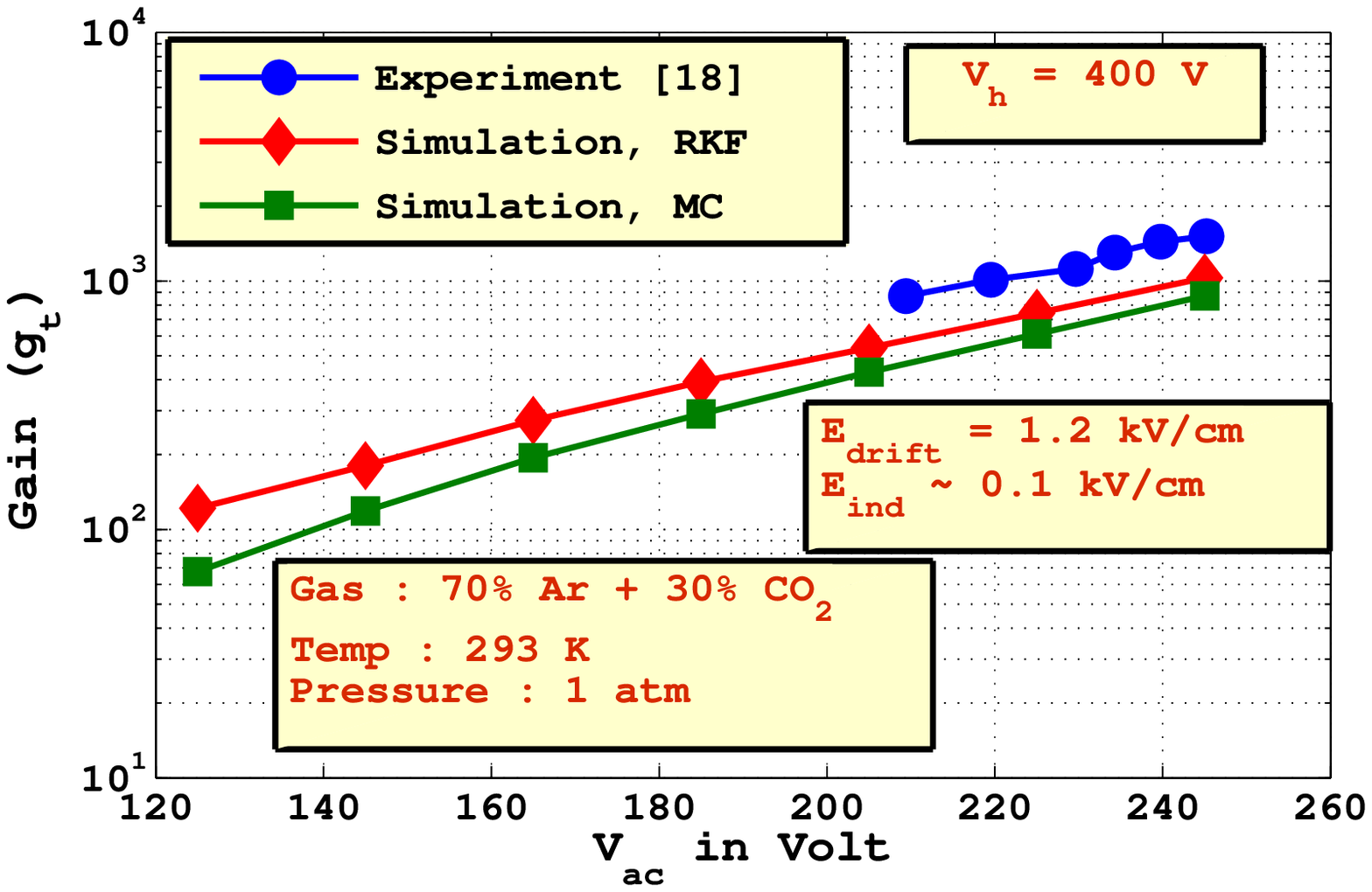}}
\caption{(a) Electron Drift line using MC Method, Variation of (b) $\epsilon_\mathrm{prim}$, 
(c) $\epsilon_\mathrm{sec}$,
(d) $\mathrm{g}_\mathrm{t}$ with $\mathrm{V}_\mathrm{ac}$ for a fixed $\mathrm{V}_\mathrm{h}$}
\label{Comp-RKF-MC}
\end{figure}

The electron trajectories using Monte-Carlo (MC) technique that takes diffusion into account is shown in figure \ref{Avalanche}.
It is expected and observed that the loss of electrons on different electrodes increases due to diffusion which naturally affects $\epsilon_\mathrm{prim}$ and $\epsilon_\mathrm{sec}$. 
Figure \ref{primary} and figure \ref{secondary} shows the variation of $\epsilon_\mathrm{prim}$ and
$\epsilon_\mathrm{sec}$ respectively with $\mathrm{V}_\mathrm{ac}$ for a fixed $\mathrm{V}_\mathrm{h}$.
A more realistic MC calculation yields a smaller efficiency than simple RKF method and thus a smaller value of gain as presented in figure \ref{gain}.  
It is to be noted that in both the RKF and MC calculations, charge induction effects of moving electrons have not been considered.
Inclusion of these effects while estimating total gain is expected to bring the MC estimates much closer to the measured values.

\section{Conclusion}
\label{sec:conclusion}

We have used Garfield+neBEM+Magboltz+Heed combination to simulate the physical processes and the performance of a MHSP detector having realistic dimensions.
A detailed study of the parametric variation of the 3D electric field, detector gain, collection efficiency has been carried out.
It has been observed that while electron focusing in a MHSP detector is almost entirely dependent on the hole voltage, its gain and electron collection efficiency requires optimum combination of both hole and strip voltages. 
The well regarded but simplistic RKF method has provided quite acceptable results in the present studies. 
However, as demonstrated, MC simulation, incorporating the effects of diffusion, promises more realistic results. 
The overall trend observed in the above studies has been found to be in agreement with the existing experimental results.
The comparative study highlights the advantages of a single MHSP detector over single GEM or MSGC.
On the one hand, while carrying out the computations we felt the necessity to have more experimental details than are available in the published literature.
On the other hand, important details such as induced component of the signal, space charge and charging up effects, estimates of manufacturing tolerances and defects have been left out of the present computations. 
In future, we hope to make progress in all these areas in order to achieve an even better understanding of these devices. 

\acknowledgments

This work has partly been performed in the framework of the RD51 Collaboration.
We happily acknowledge the help and suggestions of the members of the RD51
Collaboration. We also thank the reviewers for their valuable comments.

\end{document}